\documentclass[useAMS,usenatbib]{mn2e}
\usepackage{graphicx} 
\usepackage{times}
\usepackage{epsfig}
\usepackage{rotating}
\usepackage{sidecap}
\usepackage{longtable}
\def\lesssim{\mathrel{\hbox{\rlap{\hbox{\lower4pt\hbox{$\sim$}}}\hbox{$<$}}}}
\let\la=\lesssim
\def\gtrsim{\mathrel{\hbox{\rlap{\hbox{\lower4pt\hbox{$\sim$}}}\hbox{$>$}}}}
\let\ga=\gtrsim

\title[Low Frequency Quasi Periodic Oscillations in  GX 339-4]{Low frequency oscillations in black holes: a spectral-timing approach to  the case of GX 339-4}
\author[S. Motta et al.]{S.~Motta$^{1,2}$, T.~Mu\~noz-Darias$^{1}$, P.~Casella$^{3}$, T.~Belloni$^{1}$, J. Homan$^{4}$ \\
$^{1}$INAF-Osservatorio Astronomico di Brera, Via E. Bianchi 46, I-23807 Merate (LC), Italy\\
$^{2}$Universit\`a dell'Insubria, Via Valleggio 11, I-22100 Como, Italy \\
$^{3}$School of Physics and Astronomy, University of Southampton, Southampton, Hampshire, SO17 1BJ, UK \\
$^{4}$MIT Kavli Institute for Astrophysics and Space Research, 70 Vassar Street, Cambridge, MA 02139;
}

\begin{document}
\maketitle
\begin{abstract}
We analyzed RXTE/PCA and HEXTE data of the transient black hole binary GX 339-4, collected over a time span of eight years. We studied the properties and the behavior of low frequency quasi periodic oscillations (QPOs) as a function of the integrated broad-band variability and the spectral parameters during four outbursts (2002, 2004, 2007, 2010). 
Most of the QPOs could be classified following the ABC classification that has been proposed before. Our results show that the  ABC classification can be extended to include spectral dependencies and that the three QPO types have indeed intrinsically different properties.
In terms of the relation between QPO frequency and power-law flux, type-A and -C QPOs may follow the same relation, whereas the type-B QPOs trace out a very different relation. Type-B QPO frequencies clearly correlate with the powerlaw-flux and are connected to local increases of the count rate. The frequency of all QPOs observed in the rising phase of the 2002, 2007 and 2010 outburst correlate with the disk flux.
Our results can be interpreted within the framework of recently proposed QPO models involving Lense-Thirring precession. We suggest that type-C and -A QPOs might be connected and could be interpreted as being the result of the same phenomenon observed at different stages of the outburst evolution, while a different physical process  produces type-B QPOs.
\end{abstract}
\section{Introduction}

Quasi-periodic oscillations (QPOs) have been discovered in many systems and are thought to originate in the innermost regions of the accretion flows around stellar--mass black holes. Low-frequency QPOs (LFQPOs) with frequencies ranging from a few     mHz to $\sim$10 Hz are a common feature in almost all black hole X-ray binaries (BHB) and were already found in several sources with \emph{Ginga} and divided into different classes  (see e.g. \citealt{Miyamoto1991} for the case of GX 339-4 and \citealt{Takizawa1997} for the case of GS 1124-68). Observations performed with the Rossi X-ray Timing Explorer (RXTE) have led to an extraordinary progress in our knowledge on properties of the variability in BHBs (see \citealt{VDK2006}, \citealt{Remillard2006}, \citealt{Belloni2010} for recent reviews) and it was only after RXTE was launched that LFQPOs were detected in most observed BHBs (see \citealt{VDK2004}, \citealt{Remillard2006}). Three main types of LFQPOs, dubbed types A, B, and C, originally identified in the Power Density Spectra (PDS) of XTE J1550-564 (see \citealt{Wijnands1999}; \citealt{Homan2001}; \citealt{Remillard2002}), have been seen in several sources.  

The systematic variations in the energy spectra of transient BHBs can be identified in terms of a pattern described in an X-ray hardness-intensity diagram (HID) (see \citealt{Homan2001}; \citealt{Homan2005a}; \citealt{Belloni2005a}, \citealt{Belloni2010}). In many black hole candidates (BHC), different states are found to correspond to different branches/areas of a q-shaped HID pattern. Four main bright states (in addition to the quiescent state) have been identified in these sources, based on their spectral and timing properties (for a review \citealt{Homan2005a}; \citealt{McClintock2006}; \citealt{Belloni2010}). In particular, the analysis of the fast timing variations observed in the PDS plays a fundamental role in the state classification (see \citealt{Homan2005}, \citealt{Belloni2010}). 

Even though the general evolution and the main transitions become apparent in the HID, providing a general description of the BHB evolution, it is not enough for detailed studies. 
Many observed properties change smoothly throughout the basic diagrams (HID, see e.g. \citealt{Homan2001}; rms vs. hardness diagram, see e.g. \citealt{Belloni2010}; rms-intensity diagram, see \citealt{Munoz-Darias2011}), but some do not. It is the inspection of the fast-variability properties which indicates the presence of abrupt variations that can be taken as landmarks to separate different states. In proximity of the HIMS/SIMS transition timing properties (in particular the appearence of different types of QPOs in the PDS) constitute the sole way to distinguish between HIMS/SIMS/HSS given the absence of differences in the spectral shape. 

The different types of QPOs are currently identified on the basis of their intrinsic properties (mainly centroid frequency and width, but energy dependence and phase lags as well), of the underlying broad-band noise components (noise shape and total variability level) and of the relations among these quantities. 
Despite LFQPOs being known for several decades, their origin is still not understood and there is no consensus about their physical nature. However, the study of LFQPOs provides an indirect way to explore the accretion flow around black holes (and neutron stars). In particular, their association with specific spectral states and their phenomenology suggests that they could be a key ingredient in understanding the physical conditions that give origin to the different states. 

Several models have been proposed to explain the origin and the evolution of LFQPO in X-ray binaries.
The geometry described in \cite{Esin1997} and \cite{Done2007} allow to explain the spectral evolution seen in BHBs and forms the basis for the LFQPO model proposed by \cite{Ingram2009}, \cite{Ingram2010} and \cite{Ingram2011}, which invokes Lense-Thirring precession.  We shall refer to this model as the \emph{precession model}. 
For the case of the neutron star source 4U 1728-34, \cite{Titarchuk1999} show that LFQPO frequency is associated with radial oscillations in the boundary layer and the break frequency associated to the broad band noise in the PDS is determined by the characteristic diffusion time of the inward motion of the matter in the accretion flow. \cite{Tagger1999} associate the existence of LFQPO in X-ray binaries to an instability occurring in the inner part of disks threaded by a moderately strong vertical poloidal magnetic field.


In this paper, we examine the QPO parameters and compare them with the results of a complete spectral analysis. Since the presence and the properties of QPOs in BHBs are related to the spectral characteristics of the source, our goal is to identify the link between the spectral and fast timing variability properties and to highlight possible physical differences between the three types. The \emph{precession model} explains both the type-C QPO frequencies and the presence of the associated broad-band variability (see also \citealt{Tagger1999} for an alternative in the case of BHs) and most of spectral properties. For this reason, given the success of this  model to explain many of the observed properties we will interpret our results in the context of the  \emph{precession model} and we will attempt to extend its predictions to the three types of QPOs. 

\subsection{GX 339-4}
GX 339-4 is a Low Mass X-Ray Binary (LMXB) harboring a $> 6 M_{\odot}$ accreting black hole (\citealt{Hynes2003}; \citealt{Munoz-Darias2008}). Since its discovery (\citealt{Markert1973}), the system has undergone several outbursts, becoming one of the most studied X-ray transients. The intense monitoring carried out by RXTE during the 2002, 2004, 2007 and 2010 outburst has yielded detailed studies on the evolution of black hole states throughout the outburst (see e.g. \citealt{Belloni2005a}), making the source an ideal candidate for an extensive study on the spectral-timing properties of BHBs. Here, we use this rich data set to study the relation between the spectral and variability properties of GX 339-4 across the different outbursts.

\begin{figure}
\begin{center}
\includegraphics[width=8.5cm]{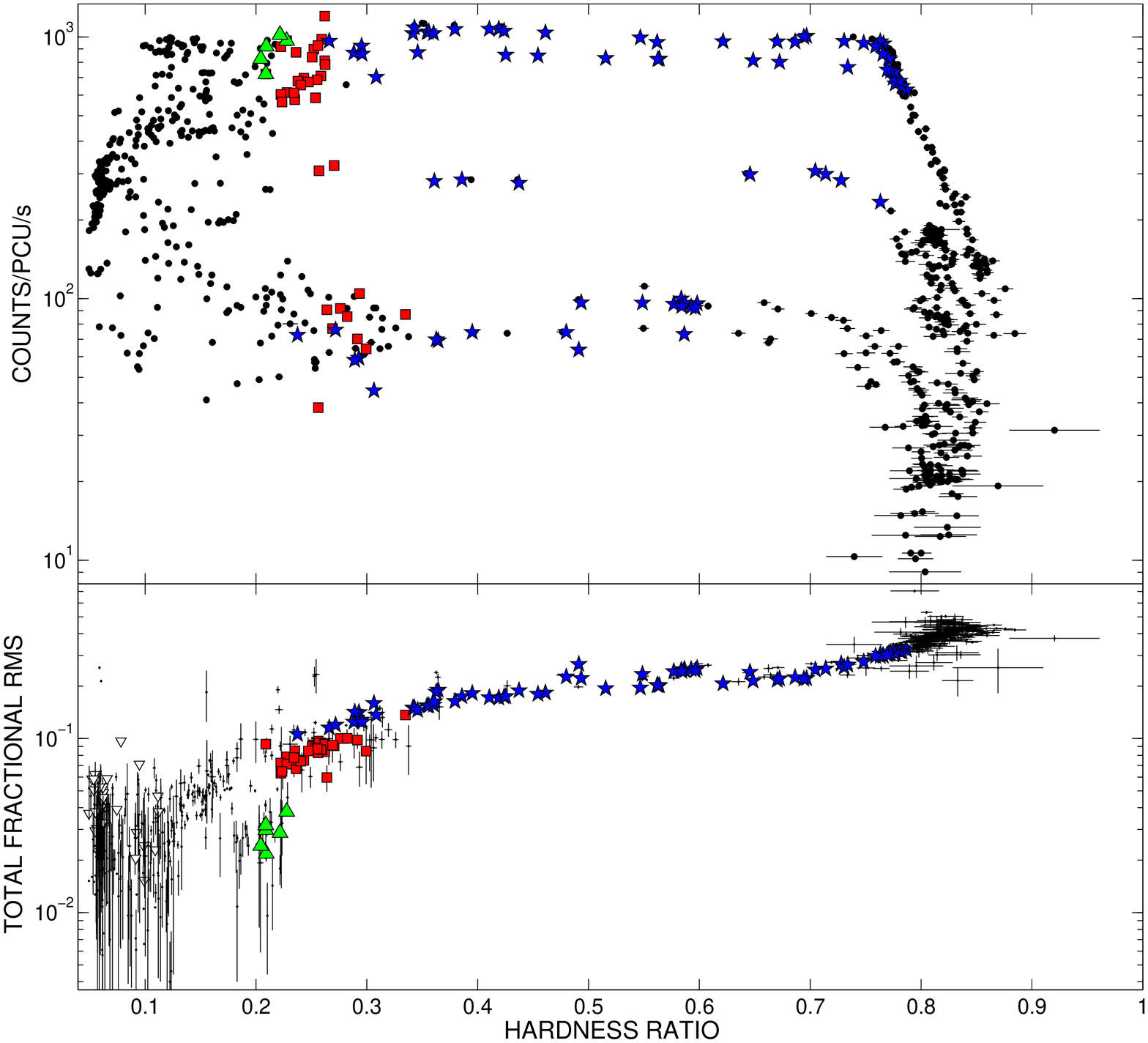}
\caption{Top panel: HID for the 2002, 2004, 2007 and 2010 outburst of GX 339-4. Each point represents a single RXTE observation. Blue stars mark Type-C QPOs, red squares mark Type-B QPOs and green triangles mark Type-A QPOs. Black dots mark all the other RXTE observations of GX 339-4 that do not show low-frequency QPOs. Bottom panel: corresponding Rms-hardness diagram. Empty triangles stand for upper limits.
}\label{fig:rms-h}
\end{center}
\end{figure}

\begin{figure}
\begin{center}
\includegraphics[width=8.5cm]{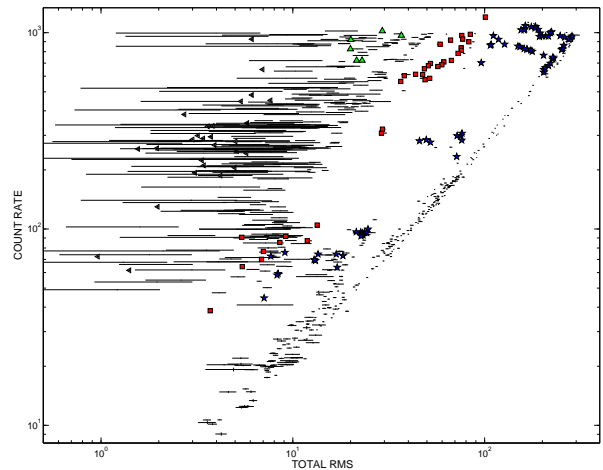}
\caption{RID for the 2002, 2004, 2007 and 2010 outburst of GX 339-4. The symbols follow the same criteria as in Fig. \ref{fig:rms-h}}\label{fig:rid}
\end{center}
\end{figure}

\section{Observations and data analysis}\label{sec:observations} 

We examined 1007 RXTE public archival observations of GX 339-4 from 2002, 2004, 2007 and 2010 outburst and selected for our analysis only observations where somewhat narrow (i.e. Q $\ge$ 2) feature was identifiable on top of peaked or power-law shaped noise components. The selection has been done by eye, then we fitted the PDS (see below) and non-significant (below 3$\sigma$) detections were excluded from the subsequent analysis. A total of 115 observations and 117 oscillations have been considered.

The RXTE data were obtained in several simultaneous modes.
{\sc Standard 2} and {\sc Standard} modes for the PCA and HEXTE instruments respectively were used to create background and dead time corrected spectra. 
We extracted energy spectra from PCA and HEXTE for each observation using the standard RXTE software within \textsc{heasoft V. 6.9}. Only Proportional Counter Unit 2 from the PCA was used since only this unit was on during all the observations. A systematic error of $0.6\%$ was added to the PCA spectra to account for residual uncertainties in the instrument calibration\footnote{http://www.universe.nasa.gov/xrays/programs/rxte/pca/doc/rmf/pcarmf-11.7 for a detailed discussion on the PCA calibration issues.}. For 2002, 2004 and 2007, we used only data coming from HEXTE/Cluster B, which was correctly working in that period. Since HEXTE/Cluster B encountered technical problems at the end of 2009 \footnote{http://heasarc.gsfc.nasa.gov/docs/xte/xhp\_new.html}, we decided to use data coming from HEXTE/Cluster A to analyze observations taken subsequently. We followed the standard procedure described in the RXTE cookbook\footnote{http://heasarc.gsfc.nasa.gov/docs/xte/recipes/hexte.html} to produce source and background spectra, using data coming from HEXTE/Cluster B to produce a preliminary background from which we derived a background spectrum to be used together with HEXTE/Cluster A spectrum.

We accumulated background corrected PCU2 rates in the Standard 2 channel bands A = 4 - 44 (3.3 - 20.20 keV), B = 4 - 10 (3.3 - 6.1 keV) and C = 11 - 20 (6.1 - 10.2 keV). A is the total count rate, while the hardness was defined as H = C/B (\citealt{Homan2005a}). 

PCA+HEXTE spectra were fitted with \textsc{XSPEC V. 11} in the energy range 4 - 40 keV\footnote{In principle it is possible to fit RXTE/PCA spectra starting from 3 keV. However, especially with sources affected by low interstellar absorption, when fitting energy spectra residuals that cannot be explained can appear. Therefore we decided to fit spectra starting from 4 keV. } and 20-200 keV respectively for data taken in 2002, 2004 and 2007 outbursts. For HEXTE spectra produced from data collected during 2010 we considered only the 20 - 150 keV spectral band. The reason for this was to exclude the harder part of the spectra that in some cases was affected by systematic problems due to an incorrect estimation of the background. For the same reason we ignored the  energy range 50 - 80 keV, where line-like residuals can be found in the spectral fits \footnote{See http://heasarc.gsfc.nasa.gov/docs/xte/whatsnew/ for details.}.   

We calculated unabsorbed fluxes for different spectral components from the best fit to the energy spectra (see Sec. \ref{sec:spectral_analysis}). We measured the total flux between 2.0 and 20.0 keV and the disk flux between 2 and 20 keV. We also measured the power-law flux in the 6 - 20 keV energy band. Since the spectral deconvolution can be problematic, we choose this energy interval to avoid contamination due to the confusion between disk and power-law component at lower energies. 
All the fluxes were normalized by the Crab flux in the respective energy bands in order to correct the fluctuations due to the variations in the instrument properties. For each outburst we used a Crab spectrum coming from an observation as close as possible to the central part of the outburst. The fluxes used for the correction are summerized in Table \ref{tab:tab_crab}.

For our timing analysis, we used {\sc GoodXenon}, {\sc Event} and {\sc Single Bit} data modes. We used custom software under \textsc{IDL} and 
for each observation we produced power density spectra (PDS) from stretches 16 seconds long in the channel band 0-35 (2-15 keV). 
We averaged the PDS and subtracted the contribution due to Poissonian noise (see \citealt{Zhang1995}). The PDS were normalized according to \cite{Leahy1983} and converted to square fractional rms (\citealt{Belloni1990}). 
The integrated fractional rms was calculated over the 0.1 - 64 Hz band. 
PDS fitting was carried out with the standard {\sc Xspec} fitting package by using a one-to-one energy-frequency conversion and a unit response. Following \cite{Belloni2002}, we fitted the noise components with three broad Lorentzian shapes, one zero-centered and other two centered at a few Hz. The QPOs were fitted with one Lorentzian each, only occasionally needing the addition of a Gaussian component to better approximate the shape of the narrow peaks and to reach values of reduced $\chi ^2 $ close to 1. 
We examined PDS in the form of dynamical power density spectra (DPDS), computing Fast Fourier Transforms of windows of data 16s long. In some cases we used shorter time windows to better follow the evolution of a narrow feature. Where transitions between different power spectral shapes were seen, we separated different time intervals in order to obtain average power spectra for each shape.
%
%
%
\begin{table} 
\renewcommand{\arraystretch}{1.3} 
\begin{center} 
\begin{tabular}{|c|c|c|c|} 
\hline 
Observation ID	&	2 - 20 keV flux	&	6 - 20 keV flux	&	Outburst	\\
            	&	erg/s/cm$^2$	&	erg/s/cm$^2$    &	Outburst	\\

\hline
\hline 
70018-01-03-00	&	3.51E-08	&	1.72E-08	&	2002		\\
90129-02-01-00	&	3.44E-08	&	1.69E-08	&	2004		\\
92802-03-06-01	&	3.40E-08	&	1.65E-08	&	2007/2010	\\
\hline
\end{tabular}
\caption{Crab fluxes used for the flux correction described in Sec. \ref{sec:observations}. Columns are: Crab observation ID, flux measured in the 2 - 20 keV band, flux measured in the 6 - 20 keV band, outburst the correction was applied to.}\label{tab:tab_crab} 
\end{center} 
\end{table} 

\subsection{The QPO classification}\label{sec:timing}


\begin{figure}
\begin{center}
\includegraphics[width=8.5cm]{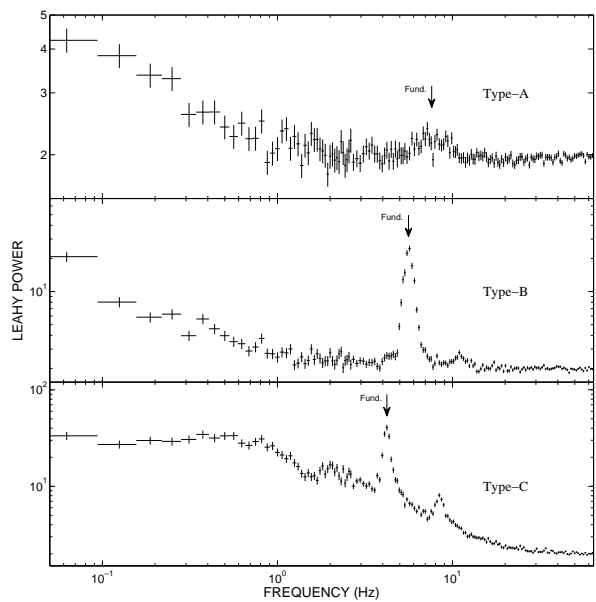}
\caption{Examples of type A, B and C QPOs from our GX 339-4 observations. The centroid peak is indicated. Upper panel: Obs. 92085--01--02--06. Middle panel: Obs. 95409-01-15-06. Bottom panel: Obs. 70109-04-01-01). The Poisson noise was not subtracted.}\label{fig:types}
\end{center}
\end{figure}

We have classified the QPOs following \cite{Casella2004}. The properties that allow one to classify the QPOs are the \textit{quality factor} ($Q = \nu_{centroid}/FWHM$) and the shape of the noise associated with the oscillation. Frequency does not allow a classification, as the characteristic frequency intervals where the three types of QPOs appear largely overlap, nor does the rms. \cite{Casella2004} and \cite{Munoz-Darias2011} quantify the noise level associated to the QPOs in terms of the total\footnote{We refer to the \emph{total fractional rms} as to the rms measured on the whole PDS. } fractional rms (0.1-64).
We could classify $\sim$98\% of the QPOs of GX 339-4 (see \ref{sec:rms_f_paragraph}). In Table \ref{tab:obs_QPO} we summarize the QPOs classification and in Fig. \ref{fig:types} we report one example for each type of QPO.

In Figure \ref{fig:rms-h} we show the HID of GX 339-4 (top panel) and the rms versus hardness diagram (bottom panel, see \citealt{Belloni2010}) including all the RXTE observations collected during 2002, 2004, 2007 and 2010 outbursts. Type-A, -B, -C QPOs are marked with green triangles, red squares and blue circles respectively. The HID is tracked counterclockwise, starting from the bottom right corner of the track. The upper and lower horizontal branches in the HID roughly correspond to the HIMS and SIMS, while the right and left vertical branches correspond respectively to the LHS and HSS. We will refer to the first part of the loop (from the right vertical branch to the left end of the top horizontal branch) as \textit{softening phase} of the outburst and to the last part (from the left end of the bottom horizontal branch back to the right vertical branch) as to the \textit{hardening phase}. 

In the following we summarize the results of the ABC classification applied to our sample.

\begin{description}

\item[\textbf{Type-C QPO - }]
In the early stages of all the four outbursts of GX339-4 (late LHS and HIMS, see Tab. \ref{tab:obs_QPO} for details), two main components can be identified in the PDS: a strong flat-top noise and one or more QPO peaks. When more than one peak is observed, the peaks are harmonically related. The strongest and narrowest peak, which is usually the central one, is taken as the fundamental. When the identification of the fundamental remained difficult because of the presence of strong harmonic peaks, we followed the evolution of the PDS shape for the selection. In Tab. \ref{tab:parameters} we report parameters for the strongest peak in the PDS.
The QPO is usually strong and narrow (Q $\ge$ 6) and the centroid frequency varies in the 0.2-9 Hz range. Only in some cases, when the oscillations are weak and appear at very low frequencies, the Q-factor is slightly lower than 10. The addition of a Gaussian component to the multi-Lorentzian model is required in a few observations in order to better approximate the peak shape. 
Type-C QPOs are observed also in the late stages of all the outbursts. We refer to  type-C QPOs observed in the lower branch as type-C$^{*}$. The PDS shows a noise component in the form of a broad Lorentzian and a QPO peak broader than at the beginning of an outburst. During the 2002 outburst, the type-C$^*$ QPOs frequencies span the 4-9 Hz range, while in the 2004 only one Type-C$^{*}$ QPO (at $\sim$ 3 Hz) is observed. In the 2007 outburst Type-C$^{*}$ QPOs are seen in a slightly lower frequency interval (2-4 Hz). A second harmonic peak is sometimes present in the PDS. 
Even though the Type-C$^{*}$ QPO centroid frequency ranges, rms properties and Q-values are different from the case of Type-C QPOs, it is possible to demonstrate that the properties of the two kinds of QPOs are continuously connected when ordered for increasing QPO frequency (see Sec. \ref{sec:result}; see also \citealt{Casella2005}).
\item[\textbf{Type-A QPO - }]
Type-A QPOs are observed in 2002 and 2004 during the SIMS, when the flux of the source is close to its maximum. The PDS show a broad QPO (Q $\le$ 3) with centroid frequency between 7.1 and 8.1 Hz associated to a weak power-law noise. Neither a subharmonic nor a second harmonic is observed. The PDS showing a Type-A QPOs have the lowest total fractional rms values of the sample. 
\item[\textbf{Type-B QPO - }]
They are observed in the SIMS and the oscillations that appear in the PDS (Q $\ge$ 6) are observable in the frequency ranges 0.8-6.4 Hz. All the type-B QPOs seen at low frequencies (i.e. below $\sim$ 3 Hz) belongs to the lower branch in the HID, while all the QPOs at higher frequencies are observed in the upper branch in the HID. 
The noise seen in the PDS is weak and the QPO peak shape is often more similar to a Gaussian rather than a Lorentzian, therefore we had to combine both components (Gaussian + Lorentzian) to obtain better fits. A weak second harmonic is often present in the PDS. Sometimes the hint of a sub-harmonic appears.

\end{description}

\subsection{Spectral Analysis}\label{sec:spectral_analysis}

In order to obtain good fits and acceptable physical parameters, a model consisting of an exponentially cut off power law spectrum reflected from neutral material (\citealt{Magdziarz1995}) was used ({\tt pexrav} in {\sc Xspec}). The reflection parameter was left free to vary, while the inclination angle was fixed at 30 degrees (notice that the results only weakly depend on the inclination value assumed). A multi-color disk-blackbody ({\tt diskbb}) was added to the model and a Gaussian emission line with centroid allowed to vary between 6.4 and 6.7 keV was further needed in order to obtain acceptable fits. The line width was constrained between 0.1 and 1.0 keV to prevent artificial broadening due to the response of XTE/PCA at 6.4 keV. 
The hydrogen column density ({\tt wabs}), was frozen to $0.5 \times 10^{22} {\rm cm}^{-2}$(\citealt{Zdziarski2004}). The addition of an iron edge, justified by the presence of the iron line, does not improve the fits significantly. 
In Table \ref{tab:parameters} we show the relevant spectral parameters for the best fits.
 
Where Type-C QPOs are detected, the photon index is seen to rise from $\sim$1.5 to $\sim$2.8 and back to $\sim$1.5 as a function of time, consistently with what previously observed (see e.g. \citealt{Motta2009}). 
Following the loop in the HID the source becomes soft while approaching the HSS and subsequently becomes hard again going back to the LHS. As a consequence the photon index increases, remains almost constant (between $\sim$2.6 and $\sim$2.8) for a while and then decreases.
When Type-A and B QPOs (during the softening phase) are seen in the PDS, the photon index is at its maximum. When type-B QPOs are observed in the hardening phase, they are associated to lower values of the photon index. Indeed, the spectra from SIMS observations in the hardening phase always show systematically lower photon indices. For a detailed analysis of the transitions between soft and hard state in GX 339-4 during 2010 outburst, see Stiele et al. (Submitted).
The photon index also correlates with the LFQPO frequency, as was already noticed by \cite{Vignarca2003} in the cases of GRS 1915+105, GRO 1655-40, XTE J1550-564, XTE J1748-288 and 4U 1630-47. The same behavior was observed in H1743-322 by \cite{MacC2009}. All type-C QPOs follow the same relation rather than several branches depending on the outburst. Type-C and type-B QPOs also overlap quite well covering the same photon index interval.

The parameters associated to the iron line and to the reflection components do not show any clear correlations with the presence of the different types of QPOs and/or particular states.

As one can see from Table \ref{tab:parameters}, not all the components of the model are present in all the observations. 
In all the observations where a Type-B or Type-A QPO is detected, a disk component ({\tt diskbb}) is visible in the spectrum. When the Type-C QPOs are detected at hardness larger than 0.6 no disk component is observed. The disk appears at hardness 0.2 - 0.6 during the softening phase. During the hardening phase a disk component was needed only in some of the spectra from 2002. 
All the other spectral components (i.e. iron line, power-law and reflection components) are always necessary to obtain good fits.

Examining the parameters related to the disk-blackbody component, it is clear that our constrains on the disc parameters are usually poor. Even when a soft component is clearly present in the spectra and is required in order to obtain good fits, it is often only marginally significant. This is expected since the working range of PCA (3-40 keV) allows to see only the high energy part of the disc black body component, above the Wien peak. It is also known that, even if the \textit{diskbb} model provides a good description of the thermal component, the derived spectral parameters should not be interpreted literally (see e.g. \citealt{Merloni2000}, \citealt{Remillard2006}). However, when the thermal component is dominant, the parameters can be taken as reliable. 

Disentangling the different spectral components could be problematic when using spectra from RXTE.  If one assumes that the hard x-ray emission comes from Comptonization of the soft disk photons on hot electrons, it is known that a simple power-law (or a power-law-like component, such as {\it pexrav}) is not appropriate for the description of the hard Comptonization tail of the spectrum at lower energies (i.e. where the hard component overlaps with the soft emission from the disk blackbody). A simple power-law does not have the low-energy cutoff that is typical of a proper comptonization model (i.e. {\it eqpair} or {\it compTT} in {\it XSPEC}) and therefore could affect the real contribution of the disk. However, when using RXTE data the adoption of a simple  power-law (or power-law-like) component is justified because the energy range where PCA spectra can be analyzed (above 3 keV) does not cover this problematic overlapping zone (just below the Wien peak of the multicolor disk-blackbody). Therefore, a simple power-law or a power-law-like model such as {\it pexrav} is appropriated for the description of PCA spectra (see \citealt{Munoz-Darias2011a} and Stiele et al. Submitted).
Despite the poor constraints on the disk parameters, the energy spectra are well fitted by the model used and the measures of the disk-fluxes reported in this work are to be considered reliable. This is supported by the fact that the disk flux correlates with the disk temperature.




\section{Results}\label{sec:result}

\subsection{Rms-frequency relation}\label{sec:rms_f_paragraph}
\begin{figure}
\begin{center}
\includegraphics[width=8.5cm]{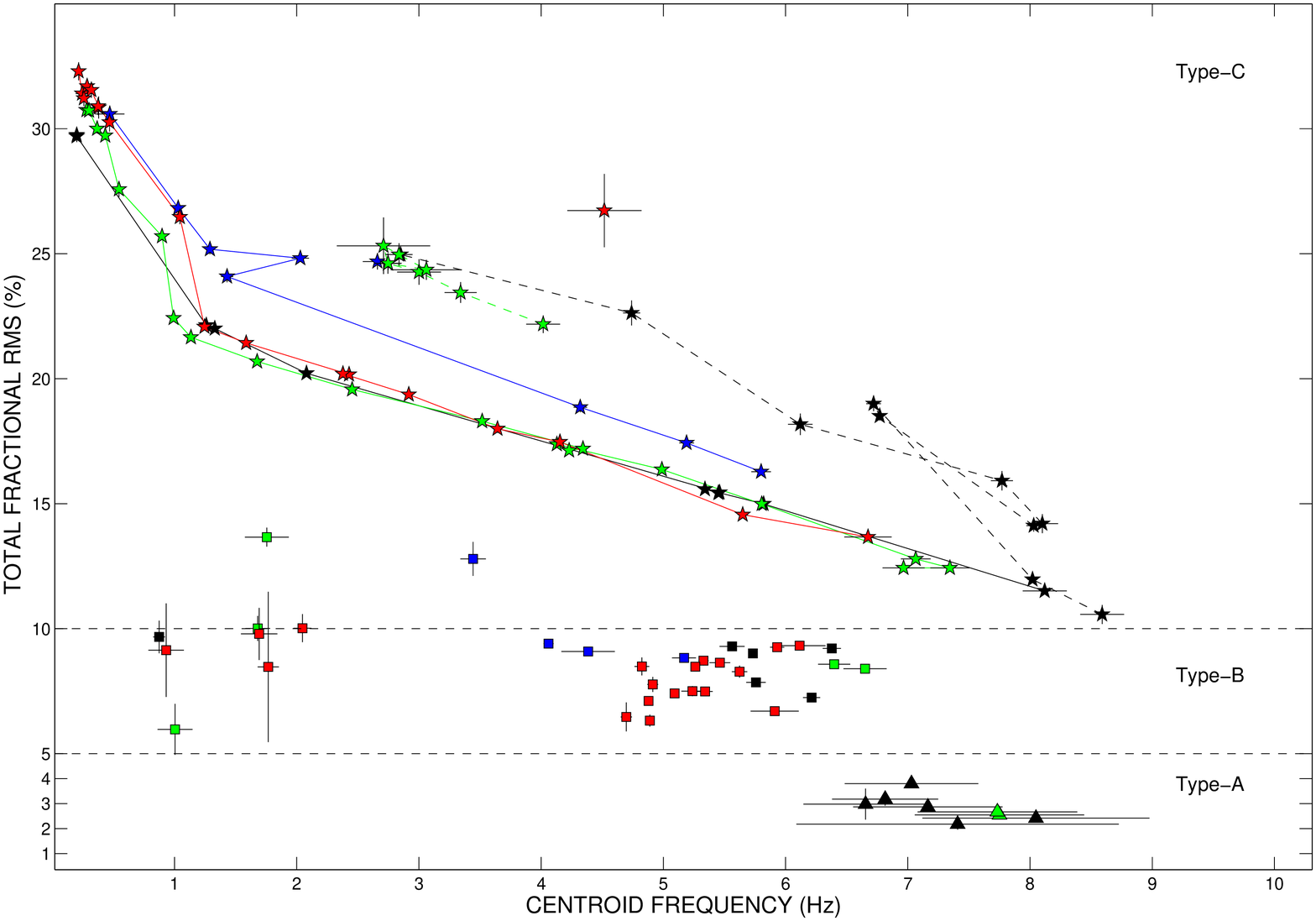}
\caption{QPO centroid frequency vs. 0.1-64 Hz fractional rms. Each point corresponds to a different observation. Symbols correspond to QPO types: circles are Type-C QPOs, triangles type-A QPOs and squares type-B QPOs. The solid lines join for each outburst type-C QPOs during the outburst softening, while dashed lines join type-C QPOs during the hardening phase. Different colors mark different outburst: black 2002, blue 2004, green 2007 and red 2010.
}\label{fig:f_rms}
\end{center}
\end{figure}


Once the QPOs of our sample were classified according to the ABC scheme, following \cite{Casella2004}, we plotted the integrated fractional rms of each PDS versus the centroid frequency (see Fig. \ref{fig:f_rms}) to probe the link between the main QPO property (the frequency) and the total variability of the source. Several groups of points, associated to the Type-A, -B, -C QPOs, can be identified. 
\begin{itemize}
\item Type-C QPOs cover the frequency range 0.1 and 9 Hz and the rms range 10-35\%. Type-C QPO frequency is clearly anti-correlated with total fractional rms. 
\item Type-A QPOs form a group at frequencies in the range $\sim$7 - 8 Hz and rms of $\sim$3\%. 
\item Type-B QPOs are located at a slightly higher rms ($\sim$ 5-10\%) in the 1-7 Hz range. 
\end{itemize}

As one can see from Figure \ref{fig:f_rms}, the softening phase (solid lines) for each outburst shows lower rms than the hardening (dashed lines). This property has already been observed in other sources (see e.g. MAXI J1659-152, \citealt{Munoz-Darias2011a}) and can probably be understood in terms of a lower disk contribution to the emission. 
Only two outliers (i.e. the  squares above and below the 5-10\% of rms in Fig. \ref{fig:f_rms}) can be identified in Fig. \ref{fig:f_rms} (Obs. \#10, \#15). Obs. \#10 shows the typical PDS shape of a type-B QPO, even though with higher rms, while Obs. \#15 shows a much more noisy PDS. Both the points lay far both from the type-C and -B region in Fig. \ref{fig:f_rms}. However, since  they follow a relation similar to type-B QPOs in a flux-frequency plot (see \ref{sec:freqflux}), we tentatively classify those QPOs as Type-B. We notice that both the observations are taken in the decay phase of the outburst (2004 and 2007 respectively).

\subsection{Frequency-Power-law flux relation}\label{sec:freqflux}

\begin{figure*}
\begin{center}
\includegraphics[width=17.5cm]{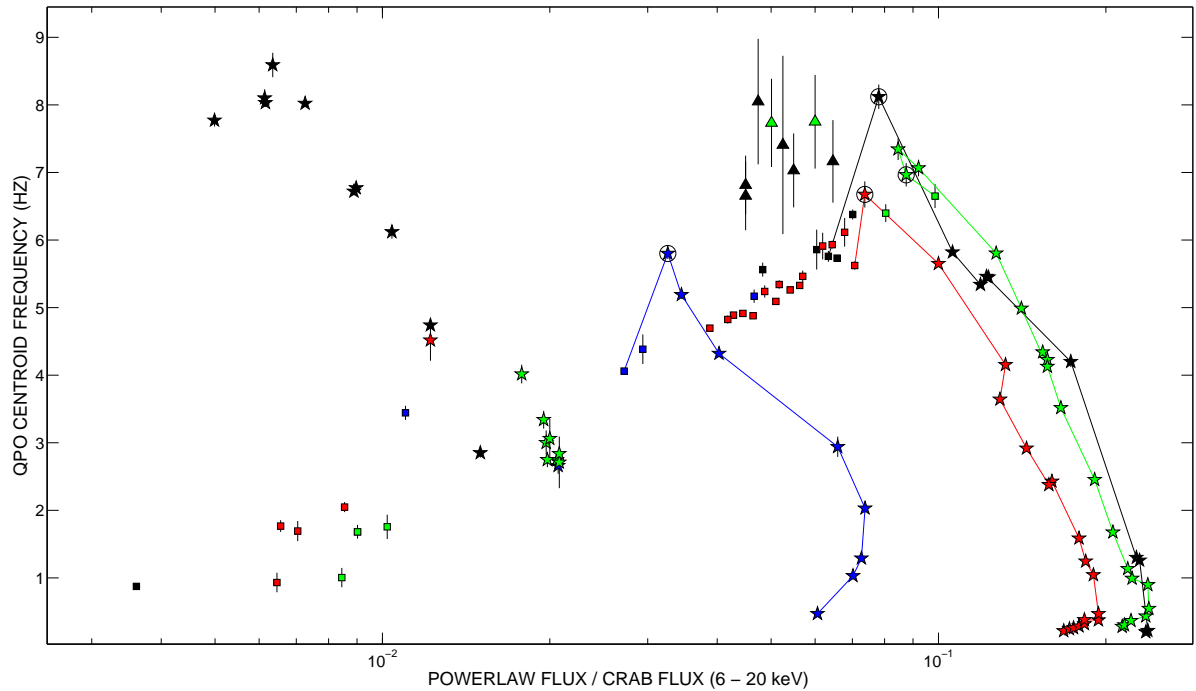}
\caption{QPO centroid frequency vs. 6-20 keV powerlaw flux, normalized to the Crab flux. Colors indicate the outburst: black stands for 2002, blue for 2004, green 2007 and red 2010. Symbols correspond to QPO types: stars are Type-C QPOs, triangles type-A QPOs and squares type-B QPOs. The solid lines join for each outburst type-C QPOs and the first type-B QPO detected after the disappearing of the Type-C QPO. The empty circles mark the type-C QPOs observed immediately before the appearance of a type-B QPO. Each point in the plot represents an entire RXTE observation in which a QPO was detected, apart from the cases in which a switch between two different types of QPOs was observed (Obs. \#1/\#26, Obs. \#4/\#27-\#28). 
}\label{fig:f_flux}
\end{center}
\end{figure*}

\begin{figure}
\begin{center}
\includegraphics[width=8.5cm]{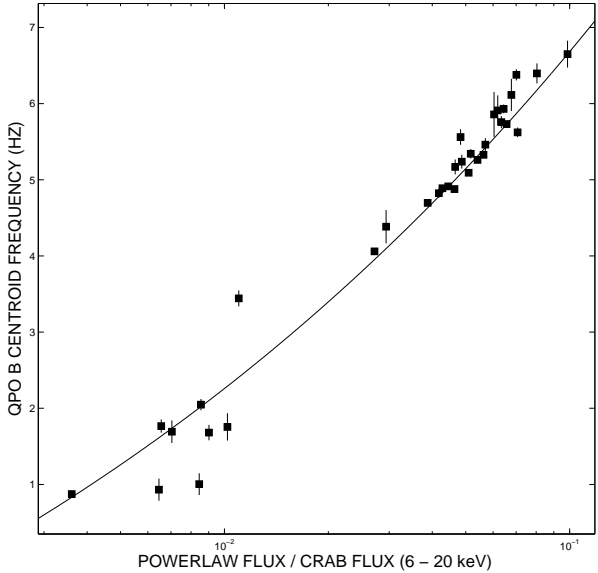}
\caption{Type-B QPOs frequency as the function of the power-law flux. The correlation is well described by a power-law of the form $y = A x^B + C$, where A = 19.4(8) , B = 0.18(6), C= -6.1(2)}\label{fig:power}
\end{center}
\end{figure}

Since LFQPOs are known to be usually  associated to the hard tail of the spectrum (see \citealt{Churazov2001} and \citealt{Sobolewska2006}, but also \citealt{Rodriguez2004} and \citealt{Rodriguez2008}, who showed tha LFQPO spectra display a moving high energy cutoff), we started  investigating the relations between LFQPO frequencies and the power-law fluxes.
The frequency-power-law flux relation is shown in Fig. \ref{fig:f_flux}.  We refer to \textit{power-law} flux as the Crab corrected flux from the power-law component in the 6 - 20 keV energy band (see Sec. \ref{sec:spectral_analysis})\footnote{The result depend only very weakly on the energy band chosen. We performed the powerlaw-flux measure also in the 2--20 keV and the getting almost identical results.}.
Different symbols represent the Type of QPOs and colors differentiate the four outbursts. For each outburst, a solid line connects in time all the QPOs from the first type-C to the first type-B QPO. The typical time interval between the last type-C and first type-B QPO is $\sim$1 day, even though sometimes other types of PDS are observed in between, especially if the transition period is not densely observed. When this happens, it might be possible to miss the appearance of a transient type-B QPO and see a type-A QPO after a type-C QPO.  In Figure \ref{fig:f_flux}, the different QPO types follow clear and separate relations as function of the hard flux.
\begin{itemize}
\item \textbf{Type-C QPOs} (stars) lie on the right part of the diagram. The points trace well-defined tracks at different flux levels for each outburst (see the solid lines in the plot). The whole 0.1 - 9 Hz frequency range is spanned. Each track roughly correspond to the late LHS and HIMS observed in the softening phase of each outburst. 
\item \textbf{Type-C$^{*}$ QPOs} follow tracks on the left part of the diagram. Those QPOs were observed in all the outbursts. Differently from what happens in the right part of the diagram, points belonging to different outburst span different and smaller frequency ranges. However, the maximum frequency reached is consistent with the softening tracks. We ascribe the fact that no QPO appear at lower frequencies to the count rate being very low. 
\item \textbf{Type-A QPOs} (triangles) cluster on a quite narrow frequency and flux range, close in frequency to the last Type-C QPOs seen before the transition to the SIMS, but at slightly lower fluxes (see Fig. \ref{fig:f_rms}). Type-A QPOs always appear in time after the detection of a type-B QPO. No Type-A QPO is observed in the left part of the plot, i.e. during the hardening phase at the end of the outburst (see also Fig. \ref{fig:rms-h}).  \ref{fig:f_rms}.
\item \textbf{Type-B QPOs } (squares) are sharply correlated with the power-law flux and the relation between frequency and powerlaw-flux is well described by a power-law of the form $y = A x^B + C$ (where A = 19.4(8) , B = 0.18(6), C= -6.1(2), see Fig. \ref{fig:power}). This correlation holds for a large range of flux, showing that these oscillations frequencies depend directly on the hard X-ray flux. However, unlike type-C QPOs, the points do not follow a clear path as a function of time.
No oscillations are seen in a given flux range in the middle of the plot. As happens for type-C and type-C$^{*}$ QPOs, this is due to the fact that also the SIMS (where type B QPOs are observed) is crossed two times, at either high or low fluxes. 

\end{itemize}

\begin{figure}
\begin{center}
\includegraphics[width=8.5cm]{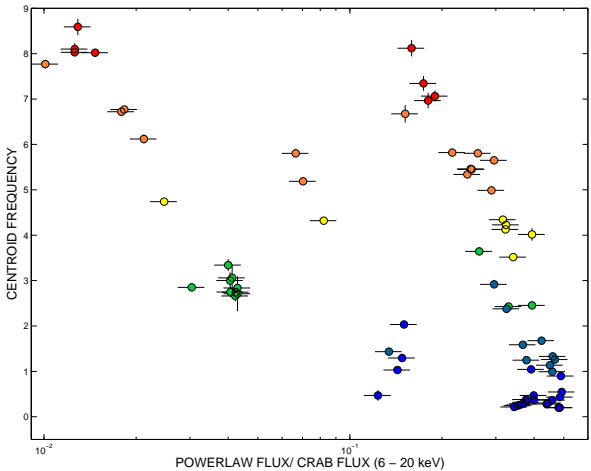}
\caption{Type-C QPO centroid frequency versus power-law flux. Different colors marks the different hardness ranges in which the QPOs where divided. From the blue to the red the ranges are: 0.7-0.8, 0.6-0.7, 0.5-0.6, 0.4-0.5, 0.3-0.4, 0.2-0.3.
}\label{fig:hard_sel}
\end{center}
\end{figure}

Since type-B QPO are found in a small hardness range, it might be argued that they behave like type-C QPOs when observed in a small hardness range. For this reason we checked whether type-C QPOs show a behavior similar to type-B QPOs once grouped in subsamples selected as a function of the hardness. We divided type-C QPOs in six subsamples and for each group we plotted the QPO centroid frequency 
as a function of the power-law flux. The result is shown in Fig. \ref{fig:hard_sel}.

Type-C QPOs' frequencies only show a weak anti-correlation with the power-law flux, especially at high frequencies. When type-C QPOs' range overlaps the hardness range where type-B QPOs are found (red points in Fig. \ref{fig:hard_sel} in the hardness range 0.2-0.3) the (weak) correlation that they follow is exactly opposite to the one shown by type-B QPOs. This fact further strengthens the difference between type-C and B QPOs.

\subsection{Frequency-Disk flux relation}\label{sec:freqflux}
\begin{figure}
\begin{center}
\includegraphics[width=8.5cm]{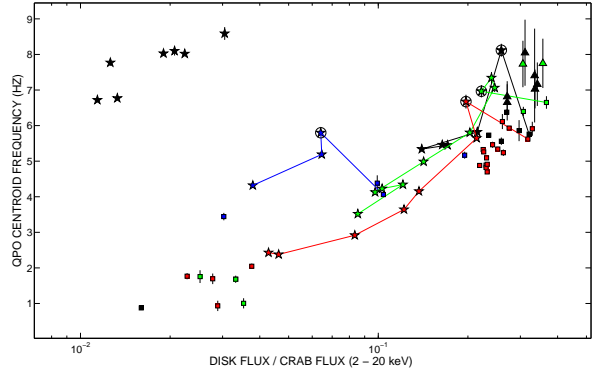}
\caption{QPO centroid frequency versus soft flux. The symbols follow the same criteria as in Fig. \ref{fig:f_flux}. 
}\label{fig:disk_f}
\end{center}
\end{figure}
In Fig. \ref{fig:disk_f} we show the relations between frequency and disk flux. Symbols and colors follow the same criteria of Fig. \ref{fig:f_flux}. We refer to \textit{disk} flux as the Crab corrected flux coming from the disk-blackbody component in the 2 - 20 keV energy range.
Since we could associate a measure of the disk flux only to a subsample of type-C QPOs, not the all of them are present in this plot.

Most of the points  trace out a well-defined track and the frequency clearly correlate with the disk flux. Those points correspond to all the QPOs seen during the upper branch of the four outbursts. Notice that for three of the four outburst (2002, 2007 and 2010 outburst) the upper branch in the HID loop is the same, while during the 2004 the upper branch is observed at a lower flux level. Note that this track includes type-C as well as type-A and B QPOs. Type-A QPOs are located in correspondence of the highest disk fluxes, while type-B QPOs cover approximately the same flux range of type-C QPOs.

We identify also a number of outliers, which correspond to two branches at different flux levels, forming other two tentative correlations. The clearest of the two is formed by all the type-C QPOs observed during the hardening phase (black points in the left upper corner of Fig. \ref{fig:disk_f}). Only during 2002 outburst it was possible to measure the disk flux during the lower branch. We ascribe this to the disk being fainter and/or colder during other outbursts than in 2002. 



\subsection{Association of Type-B QPOs with local peaks in the light curve}\label{sec:peaks}
\begin{figure}
\begin{center}
\includegraphics[width=8.5cm]{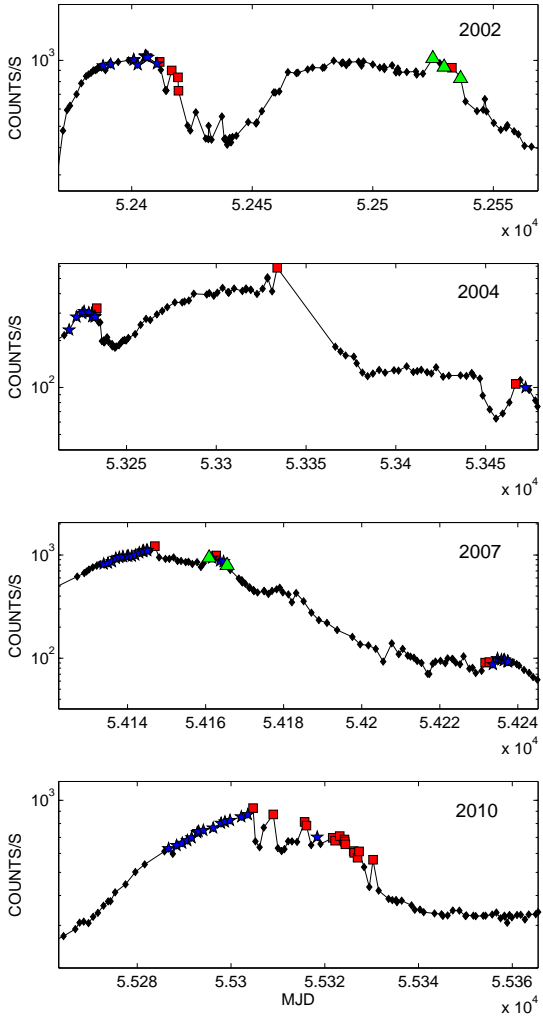}
\caption{Light curves of the 2002, 2004, 2007, 2010 outburst of GX339-4. Each point represents one entire RXTE observation. The symbols mark the different kinds of QPOs following the criteria used in Fig. \ref{fig:rms-h}. The Type-B QPOs appear associated with count-rate peaks.
}\label{fig:picchi}
\end{center}
\end{figure}

In Fig. \ref{fig:picchi} we plot sections of the light curves (PCU data, 2-20 keV) of the four outburst of GX 339-4. 
From this figure it is clear that most of the type-B QPOs are found at times of local peaks in the light curve.
In all the cases where different types of QPOs are observed within a short time interval, they follow a precise count rate segregation: type-B QPOs at highest count rates, Type-C QPOs immediately below and type-A QPOs are found at lower count rates. A relation QPO-type/count rate therefore seems to exist, although it is different from what \cite{Casella2005} observed in XTE J1859+226, where type-A QPOs are seen at highest count rates and type-B and -C appears below, even though they are not clearly separated in count rate. We notice that in all the outbursts, there is always a B at lower flux than some C in the same outburst, therefore the segregation in count rate in not absolutely true all along the outburst, but only for certain intervals (see Fig. \ref{fig:picchi}). We also notice that outbursts 2002 and 2004 had a different initial evolution in comparison to 2007 and 2010, where the count-rate peak was reached after a monotonic rise.

Despite a difference segregation in count rate, also in  XTE J1859+226 type-B QPOs are always found at hardness higher than that of type-A QPOs and lower than that of type-C QPOs. In the case of XTE J1859+226 a certain overlap was observed, while there is no overlapping in the case of GX 339-4 during a single outburst. A similar correlation between QPO types and hardness was found in XTE J1550–-564 (\cite{Homan2001}).

\subsection{Timing and spectral evolution}\label{sec:DPDS}

In four observations of our sample the PDS show rapid transitions between different shapes. In all cases the transitions involve type-B QPOs. 

\begin{figure}
\begin{center}
\includegraphics[width=8.8cm]{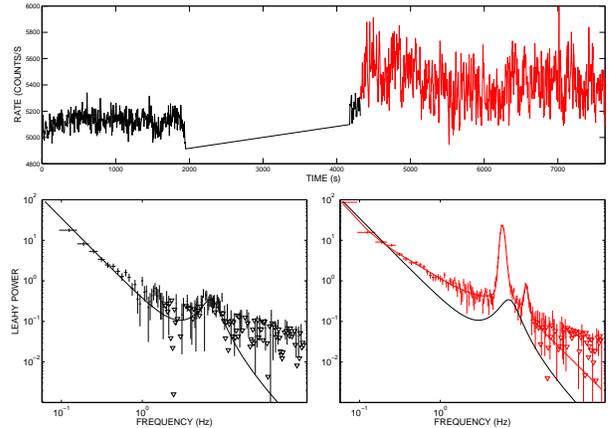}
\caption{Upper panel: Light curve for Obs. \#1/\#35. The red line marks the light curve interval where a type-A QPO was detected. The black line marks the interval where a type-B QPO was visible. Lower panel: PDS for the two time intervals \citealt{Nespoli2003}.
}\label{fig:switch1}
\end{center}
\end{figure}

\begin{figure}
\begin{center}
\includegraphics[width=8.8cm]{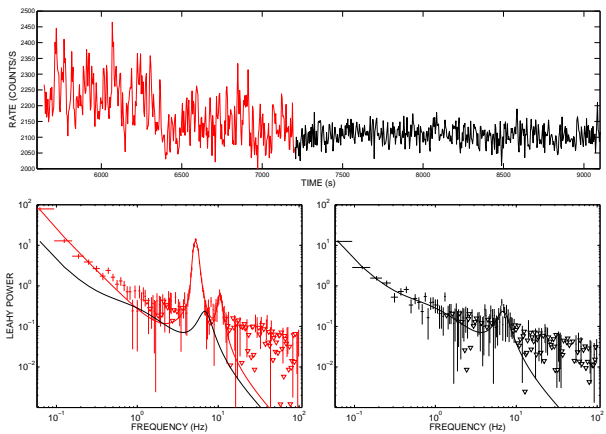}
\caption{Upper panel: Light curve for Obs. \#4/\#36-\#37 (second orbit). The red line marks the light curve interval where a type-A QPO was detected. The black line marks the interval where a type-B QPO was visible. Lower panel: PDS for the two intervals.
}\label{fig:switch2}
\end{center}
\end{figure}

In Fig. \ref{fig:switch1} and \ref{fig:switch2} we show two examples of different behaviors, for Obs. \#1/\#35 and \#4/\#36-\#37 respectively.
In the case of Obs. \#1/\#35 (Obs. ID 70109-01-07-00, Fig. \ref{fig:switch1}), a type-A QPO ($\sim$7 Hz) is present in the first part of the observation, when the observed count rate was low. In the second part the light curve shows a net increase in count rate and simultaneously the onset of a type-B QPO ($\sim$6 Hz) is observed (see \citealt{Nespoli2003} for details). At the same time, an increase of the hard flux (from $\sim$ 11\% to $\sim$13\% of the total flux) is observed, as the variation in hardness suggests (see Tab. \ref{tab:obs_QPO}). 

During Obs. \#4/\#36-\#37 (Obs. ID 70108-03-02-00, second PCA orbit, Fig. \ref{fig:switch2}) a similar situation can be observed. In the first orbit the PDS shows a type-A QPO ($\sim$7 Hz). In the second orbit the source count rate dropped abruptly from $\sim$2200 counts/s to $\sim$2100 counts/s\footnote{Note that the counts are taken from \it{EVENT} mode data and they are not normalized per number of PCUs on.} in few seconds. A type-B QPO ($\sim$5.6 Hz), that was visible in the first part of the observation disappears leaving place to a type-A QPO ($\sim$7 Hz), observable until the end of the interval and during the complete third orbit. Analogous to the previous case, a variation in flux takes place. When the type-B QPO disappears, the power-law flux is seen to decrease abruptly (from $\sim$ 15\% to $\sim$11\% of the total flux in $\sim$1s). The frequency of the type-A QPO before the appearance of the type-B QPO and and after its disappearance in the light curve  is the same.

Two other cases are found in Obs. \#5 and \#30 (Obs. ID 70110-01-47-00 and 95409-01-19-00), where, in correspondence of a rise in the count rate, a type-B QPO takes the place of power-law-shaped noise (i.e. no type-A or C QPO is observable before the onset of the type-B QPO). In all the mentioned cases it is clear that spectral differences can be very subtle, much more than the timing changes.

Similar fast transitions between different types of QPOs have already been observed in GX 339-4 (\citealt{Miyamoto1991}) and in other sources, such as XTE J1859+226 (\citealt{Casella2004}) and GS 1124-68 (\citealt{Takizawa1997}). For XTE J1859+226, the type-B QPO seems to be associated to a flaring behavior. However, also in this source the type-A QPO is always seen at slightly higher frequencies than the type-B QPOs.

As it happens for type-A/B QPOs, direct switching from/to type-C/type-B can be observed in GX339-4 as in other sources (see e.g. \citealt{Miyamoto1991a}, \citealt{Takizawa1997}). However, the switch between type-B and -C QPOs is not as sharp as in the case of type-B and type-A QPOs: the transition between the two types usually comes with a complex behavior and Type-B QPOs appears in correspondence to peaks in the light curve (occurring at timescales of few seconds), consistently with what described in Sec. \ref{sec:peaks}, and type-C QPOs are seen where the count rate drops. For this reason, transitions between/from type-C and type-B QPOs are not easily detectable and are worthy of a more detailed analysis that is beyond the scope of this work. For a more detailed study, see Homan et al. (in prep). 


\section{Discussion}\label{sec:discussion}

We analyzed RXTE/PCA and HEXTE data collected over eight years of observations of the transient BHB GX 339-4 to study the properties and the behavior of LFQPOs. 115 out of 117 oscillations could be classified into the three main types (A, B, C). The coherent scenario we constructed can be compared to that of other systems (such as XTE 1859+226 and XTE J1550-564) for which the ABC classification has been performed. 
Different properties and relations emerge from the analysis, allowing a better characterization of the different types of QPOs. Our results confirm that the ABC classification can be extended to include spectral dependencies. The main parameters of the different types of LFQPOs observed in GX 339-4 are summarized in Table \ref{tab:tab_ABC}. 

\begin{table} 
\renewcommand{\arraystretch}{1.3} 
\begin{center} 
\begin{tabular}{|c|c|c|c|} 
\hline 
		& A 		& B 			& C 				\\
\hline
\hline 
$\nu$ 	& 6.5-8 Hz 	& 0.8-6.4 Hz 	& 0.2-9 Hz 			\\
Q 		& 1-3 		& $\geq 6 (\geq 2)^*$ 		& $\geq 6 (\geq 2)^*$ 		\\
rms 	& $\leq5\%$ & $5-10\%$ 		& $\geq10\%$ 		\\
noise 	& weak red  & weak red 		& strong flat-top 	\\

\hline
\end{tabular}
\caption{Summary of type-A, -B and -C QPOs properties in GX 339-4.
(*) The bracket values correspond to the hardening phases.}\label{tab:tab_ABC} 
\end{center} 
\end{table} 

\begin{description}
\item[\textbf{Type-C QPO:}]this oscillation is found in LHS and HIMS (see \citealt{Belloni2010} and Homan et al 2011 in prep.). It is observed in both the softening and hardening outburst intervals, even though during the hardening it results weaker. It appears at hardness values in the 0.2 - 0.8 range and over a large frequency range (0.2 - 9 Hz). Its centroid frequency rises as function of the time as the source undergoes the softening phase and decreases as the source hardens during the decay phase. All type-C QPOs are spread above a variability level below which only type-B and type-A QPOs are observed (see \citealt{Casella2004} and \citealt{Munoz-Darias2011}). 
The observed 0.1-64 Hz rms has values between 10 and 35\%. Those are positively correlated with the hardness ratio (see e.g. \citealt{Belloni2010}) and negatively correlated to the frequency (see Fig. \ref{fig:f_rms}).
Type-C QPOs form a clear but complex pattern in a frequency versus power-law-flux plane. The frequency type-C QPOs correlates with the disk flux and form different branches in a frequency versus disk-flux plane, corresponding to the softening and hardening phases. Type-C QPO frequencies correlates well with the hardness, stressing a clear dependence on the spectral shape.\\
\item[\textbf{Type-A QPO:}] this QPO is usually observed in the SIMS, which is indeed defined on the basis of the appearance of type-A and -B QPOs. It is found in a narrow hardness (0.20 - 0.22), frequency (6.5 - 8.0 Hz) and rms ($\sim$ 2 - 3\%) range. The frequency at which they are found is always very close to the frequency of the last type-C QPOs observed before the transition to the SIMS. We observed type-A QPO only during the softening phase (i.e. along the upper horizontal branch in HID). The lack of this type of QPO during the hardening phase can however be ascribed to the lower statistics, as the feature is weak and broad.
This is the QPO type that is found to be associated to the lowest total fractional rms in our sample. In a frequency versus power-law-flux plot, type-A QPOs appear to be grouped close to the high-frequency end of the tracks defined by the type-C QPOs and are found around the same frequency of the type-C QPOs observed close to the transition to the SIMS.\\
\item[\textbf{Type-B QPO:}]its presence defines the SIMS. The frequency range where it is observed is 0.8 - 6.4 Hz. In addition, all type-B QPOs are observed at lower frequencies with respect to the type-C QPOs observed just before the transition to the SIMS. As for type-C QPOs, these oscillations are seen in both the softening and hardening outburst phase. Total fractional rms and hardness values are lower than in the case of type-C, but higher than for type-A QPOs, ranging in the intervals 5 - 10\% and 0.2 - 0.3 respectively. It is noticeable that between 5 and 10\% broad band total fractional rms only type-B QPOs are observed (see \citealt{Munoz-Darias2011}). This type of QPO follows a sharp frequency-flux correlation for both disk and powerlaw flux (Fig. \ref{fig:f_flux} and \ref{fig:disk_f}). 

From Fig. \ref{fig:f_flux} one can see that type-B QPOs follow a clearly different path compared to the other classes, suggesting the presence of an intrinsic difference. 
\end{description}

It is interesting to compare some of the properties reported here for GX 339-4 with those observed in other sources. \citealt{MacC2009} performed a detailed spectral analysis and a supporting timing analysis on the BHC H1743-322 and compared their results with the BHC XTE J1550-564, while \cite{Sobczak2000} compared the properties of XTE J1550-564 with GRO J1655-40. These three sources showed a behavior similar to GX 339-4 in the frequency-disk flux plane, spanning a larger flux range. 
\cite{Sobczak2000} also investigated the relations between frequency and powerlaw flux, finding opposite relations for GRO J1655-40 and XTE J1550-564. The frequency/powerlaw-flux correlation in GX 339-4 is similar to that of GRO J1655-40, but  the frequency range covered by the two sources do not overlap, making impossible a direct comparison. 

The relation between QPO  frequency and power-law photon index is the same for GX 339-4 and H1743-322, XTE J1550-564 and GRO J1655-40 (\citealt{Sobczak2000}; \citealt{MacC2009}), following the original correlation observed by \cite{Vignarca2003}. 
As for the frequency-total rms plane, which was originally presented for XTE J1859+226 (\cite{Casella2005}), the same relation is present also in H1743-322 and XTE J1550-564 (\citealt{MacC2009})
.

\subsection{Similarities and differences: a common origin for QPO-types?}\label{sec:similarities}
The different types of QPOs often show similar/compatible properties (eg. centroid frequency range, QPO profile, quality factor values), some of which suggest that there could be a common origin for the different classes. However, there are systematic differences that cannot be ignored.

As is clear from Fig. \ref{fig:f_rms}, a stringent relation between the total (0.1 - 64 Hz) fractional rms values and QPO-type exists and the different types of QPOs correspond to different and well separated rms ranges (see \citealt{Casella2004} and \citealt{Munoz-Darias2011}). At the same frequency two or sometimes even three different types of QPOs can be seen (not simultaneously) depending on the variability level at which the system is observed.

Despite the clear separation in rms, type-C and type-A QPOs follow a similar hard-flux/frequency relation. In addition, type-A QPOs and the type-C QPOs observed just before the HIMS/SIMS transition show very similar frequencies (see \ref{fig:f_rms} and  Tab. \ref{tab:obs_QPO}), while type-B QPOs are systematically found at lower frequencies. Unfortunately type-A QPOs constitute only a small part of our entire sample of QPOs (8 out of 117), therefore it is not possible to completely exclude a bias in the frequency association due to the small number of detections.

It is clear that Type-A and -C QPOs are significantly different timing features, in particular for what concerns the broad band noise associated to the two types of oscillations (strong broad band noise in the case of type-C QPOs and weak power-law or weak peaked noise in the case of type-A QPOs). However, this is not enough to rule out the possibility that type-A and type-C QPOs share a common physical origin.

Here we discuss the evolution of the LFQPOs in the framework of the model proposed by \cite{Done2007}, that also suggest possible explanation of the simultaneous spectral transition seen during an outburst. In the geometry that these authors assume, the outer accretion flow takes the form of a cool, geometrically thin, optically thick accretion disk truncated at some radius, which is larger than the last stable orbit (in the HIMS). The inner accretion flow, instead, forms a hot, geometrically thick, optically thin configuration. The inward movement of the truncation radius with increasing mass accretion rates within the hard states gives a physical basis for the hard-soft transition when the disk finally reaches the last stable orbit. The inner disk radius evolution gives also a possible origin for the LFQPOs and to the associated noise observed in the power density spectra. In this scenario LFQPOs (and in particular type-C QPOs) arise from the vertical Lense-Thirring precession (\citealt{Stella1998}) of the misaligned inner hot flow, while the broad band noise arises from propagation of Magneto Rotational Instability (MRI, \citealt{Balbus1991} \footnote{Notice that MRI is not able to produce the jets observed during the hards states, therefore a different/additional mechanism must be also at work during the hard state. This is also valid during the transition to the SIMS where relativistic ejections can be observed. } ) fluctuations of the same hot flow (see \citealt{Ingram2009}, \citealt{Ingram2010}, \citealt{Ingram2011} for details)\footnote{For alternative models on the origin of LFQPOs see e.g.  \cite{Wagoner2001}, \footnote{Tagger1999}, \cite{Titarchuk1999}, \citealt{Tagger1999}.}. The QPO frequency depends on the truncation radius and also on the optical depth of the inner hot flow that produces them, as well on the inclination of the system (see also \citealt{Homan2005b}) and to the optical depth of the precessing inner hot flow.
In this scenario, the fact that type-A QPOs and the type-C QPOs observed close to the transition are seen at the same frequency, suggest either that they are produced in a region at similar radii or in a medium at the same physical conditions (i.e. same optical depth). This second possibility is supported by the clear difference in broad-band noise level in the PDS, that is thought to be related to MRI. The type-C QPO width sets the typical timescale for the QPO signal to remain coherent. Since the QPO is not always present in the light curve, there also exists an excitation timescale responsible for the QPO appearance (\citealt{Lachowicz2010}). The QPO can be broadened if the excitation timescale becomes shorter than the QPO timescale, as the excitation presumably interrupts the coherent QPO light curve by phase randomizing it. In the framework set by the global Lense-Thirring precession of the hot flow  model (\citealt{Fragile2007}; \citealt{Ingram2009}; \citealt{Ingram2011}) the QPO could be triggered by turbulence, resulting in a vertical kick applied by the bending waves propagating in the outer accretion flow to the precessing inner hot flow at random phase. If more than one kick occur within the QPO timescale, the QPO coherent light curve is interrupted by a random phase shift. As a result, the type-C QPO would evolve in a broader and fainter feature (Chris Done, private communication). Thus, type-A and -C QPOs could be the result of the same physical process, i.e. Lense-Thirring precession. However, given the undeniable link between the type-A QPOs and very weak noise, it remains a fact that whatever process broadens and weakens the QPOs, should be also responsible for the collapse of the noise.

\subsection{The peculiar case of the type-B QPO}

Type-B QPOs show properties that differentiate them from the other two classes. 
Analyzing the frequency-hard flux relation and the frequency-rms relation, it is evident that there is a clear discontinuity within the pattern defined by type-B QPOs and other types of QPOs. While all the QPO frequencies seem to correlate with the soft flux, only type-B QPOs show a sharp correlations with the hard flux. This is remarkable because when type-B and type-A/-C QPOs are seen at similar hardness (type-C QPOs observed just before the transition and all type-A QPOs), there are no differences in the spectral shape, as one might deduce from the HID, but only in flux (see below).
In addition, type-B QPOs are systematically found at lower frequencies with respect to the last type-C QPOs and type-A QPOs (see \ref{fig:f_rms} and \ref{fig:f_flux}). 

Type-B QPOs can be transient (i.e appear/disappear in few seconds and are observable only for short periods, see also \citealt{Takizawa1997}) and vary significantly around their centroid frequency, with a characteristic time scale of $\sim$10s (see \citealt{Nespoli2003}). Also type-C QPOs can appear and disappear in few seconds, but they remain observable for long periods and can be easily followed in their frequency evolution during the hard-to-soft or soft-to-hard transition. Because of its intrinsic faintness, type-A QPOs cannot be followed as can be done for type-B and -C QPOs, (see \citealt{Nespoli2003}). 

A noticeable peculiarity of type-B QPOs is the association to flux peaks. This is particularly evident for a direct switch from or to a type-B QPO (see Sec. \ref{sec:DPDS}). The association of type-B QPOs with increases in the count rate can be seen both in the total light curve  (i.e. peaks in the count rate observed in the total light curve of the source, see also \citealt{Fender2009}) and on shorter timescales (i.e. when sudden increases in count rate take place during a single RXTE pointing). Under the assumption that the count rate tracks the accretion rate, type-B QPOs would be related to increases of the local accretion rate. Alternatively, the increase in the count rate associated to those QPOs could be related to the presence of a jet component that would contribute to the hard emission. A third possibility is that type-B QPOs might occur simultaneously to sudden changes in the geometry or radiative efficiency, which would possibly cause variations in the relative contribution of the emitting component to the spectrum and in the flux.  

Starting from the \emph{precession model} and following a reasoning similar to that described above, we argue that type-B QPOs are either produced in a different region located at larger radii (with respect to the region where the type-C QPOs close to the transition and type-A QPOs might originate, to match the lower frequencies observed) or coming from a modulation operated by a medium with different physical properties. In the first case, it is necessary to find a process different from the vertical Lense-Thirring precession, that would be able to produce modulations at larger radii. This hypothesis is supported by the fact that when fast switches from/to type-A/-B QPOs are observed, the type-A QPOs is consistent with being still present when the type-B QPOs appears (see also \citealt{Nespoli2003}). This is also valid for type-A and -B detected in separated observations.  Such a property suggests that two different, eventually simultaneous mechanisms, might be responsible of the production of type-C/A QPOs and type-B QPOs. 
For the second case, a process able to trigger fast transitions in the physical properties of the plasma would be needed. This hypothesis is supported, as for type-A QPOs, by the fact that the transition to type-B QPOs is associated to the significant difference in the PDS broad band noise level.  
There is also a third possibility: type-B QPOs could be produced in the same region where type-C QPOs come from, but thanks to a different pecession mode. In the \emph{precession model} the QPO arises from the surface density-weighed Lense-Thirring-precession over the inner hot flow. A sudden change in the surface density profile - for example due to a jet ejection from the very inner regions of the accretion flow -  would result in a different weighing and also a different precession frequency.


\section{Conclusions}
The large amount of RXTE observations of GX 339-4 in the past eight years allowed us to analyze the spectral and temporal behavior of the source over four outbursts. We considered all the observations where a low frequency oscillation was observed and performed a complete spectral and timing analysis. Almost all the oscillations observed in the PDS could be classified following the ABC classification and the three types of QPOs display different dependences on the spectral and timing parameters, further strengthening their  intrinsic differences.

We conclude that type-B QPOs show properties that clearly differentiate them from other types of QPOs. Their frequencies  clearly correlate with the powerlaw flux, tracing a complete different pattern with the respect to type-A and -C QPOs. Type-B QPOs follow a different behavior also in a rms vs rms plane. In addition, they show a peculiar association to increases in count rate that could reflect changes in accretion rate and/or geometry in the system.

All the types of QPOs can be explained through the \emph{precession model} (\citealt{Ingram2009}, \citealt{Ingram2010}, \citealt{Ingram2011}) as the result of the vertical Lense-Thirring precession of a optical translucent inner hot flow in a truncated disk geometry. However, the characteristic properties of type-B QPOs suggest that they could be the effect of a physical phenomenon different from the Lense-Thirring precession and possibly somehow related to the transition/jet ejection mechanism.

\vspace{1cm}
\noindent  SM acknowledges Chris Done for hospitality during her visit at University of Durham and for useful discussions that led to significant improvement of this work and the referee Jerome Rodriguez for useful comments and suggestions.
SM and TB acknowledge support from grant ASI-INAF I/009/10/. TMD acknowledges Univeristy of Amsterdam and Southampton for hospitality during his visits.
The research leading to these results has received funding from PRIN INAF 2007 and from the European Community's Seventh Framework Programme (FP7/2007-2013) under grant agreement number ITN 215212 \textquotedblleft Black Hole Universe\textquotedblright. PC acknowledges support from a EU Marie Curie Intra-European Fellowship within the 7th European Community Framework Programme, under contract no. 2009-237722.  This work has been partially funded by the Spanish MEC under the Consolider-Ingenio 2010 Program grant CSD2006-00070: ‘First Science with the GTC’ (http://www.iac.es/consolider-ingenio-gtc/).

\newpage

\onecolumn

\begin{center} 
\begin{longtable}{|c|c|c|c|c|c|c|c|c|} 
\caption[]{Power-spectral classification and variability parameters. Only observations with evidence of low frequency QPOs are listed.}\label{tab:obs_QPO}\\
\endfirsthead

\multicolumn{9}{c}%
{{\tablename\ \thetable{} -- continued from previous page}} \\ \hline
$\#$				&	Obs. ID	&		MJD		&	Outburst	&		Hardness ratio				&		Total fractional rms				&		QPO centroid Frequency				&	QPO type	&	State	\\
\hline																																
\hline	
\endhead

\hline \multicolumn{9}{c}{{Continued on next page}} \\
\endfoot

\hline 
\endlastfoot
\hline																																		
																																		
$\#$				&	Obs. ID	&		MJD		&	Outburst	&		Hardness ratio				&		Total fractional rms	&		QPO centroid Frequency				&	QPO type	&	State	\\
\hline																																		
\hline																																			

	$	1	$	&	70109-01-07-00b	&	$	52411.601	$	&	2002	&	$	0.228	\pm	0.001	$	&	$	7.9	\pm	0.1	$	&	$	5.8	\pm	0.1	$	&	B	&	SIMS	\\
	$	2	$	&	70110-01-14-00	&	$	52416.596	$	&	2002	&	$	0.252	\pm	0.001	$	&	$	9.2	\pm	0.1	$	&	$	6.4	\pm	0.1	$	&	B	&	SIMS	\\
	$	3	$	&	70110-01-15-00	&	$	52419.238	$	&	2002	&	$	0.251	\pm	0.001	$	&	$	9.0	\pm	0.1	$	&	$	5.7	\pm	0.03	$	&	B	&	SIMS	\\
	$	4	$	&	70108-03-02-00b	&	$	52419.432	$	&	2002	&	$	0.209	\pm	0.001	$	&	$	9.3	\pm	0.2	$	&	$	5.6	\pm	0.1	$	&	B	&	SIMS	\\
	$	5	$	&	70110-01-47-00b	&	$	52532.749	$	&	2002	&	$	0.222	\pm	0.001	$	&	$	7.2	\pm	0.2	$	&	$	6.2	\pm	0.1	$	&	B	&	SIMS	\\
	$	6	$	&	70110-01-89-00	&	$	52707.915	$	&	2002	&	$	0.256	\pm	0.004	$	&	$	9.7	\pm	0.7	$	&	$	0.9	\pm	0.05	$	&	B	&	SIMS	\\
	$	7	$	&	90110-02-01-03	&	$	53232.993	$	&	2004	&	$	0.257	\pm	0.001	$	&	$	9.4	\pm	0.2	$	&	$	4.1	\pm	0.04	$	&	B	&	SIMS	\\
	$	8	$	&	90704-01-02-00	&	$	53233.389	$	&	2004	&	$	0.271	\pm	0.001	$	&	$	9.1	\pm	0.2	$	&	$	4.4	\pm	0.2	$	&	B	&	SIMS	\\
	$	9	$	&	60705-01-84-02	&	$	53333.899	$	&	2004	&	$	0.254	\pm	0.001	$	&	$	8.8	\pm	0.1	$	&	$	5.2	\pm	0.1	$	&	B	&	SIMS	\\
	$	10	$	&	91105-04-10-00	&	$	53466.753	$	&	2004	&	$	0.293	\pm	0.003	$	&	$	12.8	\pm	0.7	$	&	$	3.4	\pm	0.1	$	&	B?	&	SIMS	\\
	$	11	$	&	92035-01-04-00	&	$	54147.011	$	&	2007	&	$	0.262	\pm	0.001	$	&	$	8.4	\pm	0.1	$	&	$	6.7	\pm	0.2	$	&	B	&	SIMS	\\
	$	12	$	&	92085-01-03-01	&	$	54162.665	$	&	2007	&	$	0.259	\pm	0.001	$	&	$	8.6	\pm	0.1	$	&	$	6.4	\pm	0.1	$	&	B	&	SIMS	\\
	$	13	$	&	92704-03-10-00	&	$	54231.604	$	&	2007	&	$	0.264	\pm	0.002	$	&	$	6	\pm	1	$	&	$	1.0	\pm	0.1	$	&	B	&	SIMS	\\
	$	14	$	&	92704-03-10-11	&	$	54232.658	$	&	2007	&	$	0.276	\pm	0.002	$	&	$	10.0	\pm	0.5	$	&	$	1.7	\pm	0.1	$	&	B	&	SIMS	\\
	$	15	$	&	92704-03-10-12	&	$	54233.565	$	&	2007	&	$	0.335	\pm	0.002	$	&	$	13.7	\pm	0.4	$	&	$	1.8	\pm	0.2	$	&	B?	&	SIMS	\\
	$	16	$	&	95409-01-15-02	&	$	55304.714	$	&	2010	&	$	0.256	\pm	0.001	$	&	$	8.3	\pm	0.2	$	&	$	5.6	\pm	0.1	$	&	B	&	SIMS	\\
	$	17	$	&	95409-01-15-06	&	$	55308.983	$	&	2010	&	$	0.236	\pm	0.001	$	&	$	6.7	\pm	0.1	$	&	$	5.9	\pm	0.2	$	&	B	&	SIMS	\\
	$	18	$	&	95409-01-16-05	&	$	55315.695	$	&	2010	&	$	0.262	\pm	0.001	$	&	$	9.3	\pm	0.1	$	&	$	6.1	\pm	0.2	$	&	B	&	SIMS	\\
	$	19	$	&	95409-01-17-00	&	$	55316.114	$	&	2010	&	$	0.262	\pm	0.001	$	&	$	9.3	\pm	0.1	$	&	$	5.9	\pm	0.1	$	&	B	&	SIMS	\\
	$	20	$	&	95409-01-17-05	&	$	55321.718	$	&	2010	&	$	0.243	\pm	0.001	$	&	$	7.5	\pm	0.1	$	&	$	5.3	\pm	0.1	$	&	B	&	SIMS	\\
	$	21	$	&	95409-01-17-06	&	$	55322.230	$	&	2010	&	$	0.238	\pm	0.001	$	&	$	7.5	\pm	0.2	$	&	$	5.2	\pm	0.1	$	&	B	&	SIMS	\\
	$	22	$	&	95409-01-18-00	&	$	55323.210	$	&	2010	&	$	0.259	\pm	0.001	$	&	$	8.6	\pm	0.1	$	&	$	5.5	\pm	0.1	$	&	B	&	SIMS	\\
	$	23	$	&	95335-01-01-07	&	$	55324.189	$	&	2010	&	$	0.255	\pm	0.001	$	&	$	8.7	\pm	0.2	$	&	$	5.3	\pm	0.04	$	&	B	&	SIMS	\\
	$	24	$	&	95335-01-01-00	&	$	55324.254	$	&	2010	&	$	0.247	\pm	0.001	$	&	$	8.5	\pm	0.1	$	&	$	5.3	\pm	0.03	$	&	B	&	SIMS	\\
	$	25	$	&	95335-01-01-01	&	$	55324.393	$	&	2010	&	$	0.240	\pm	0.001	$	&	$	7.4	\pm	0.1	$	&	$	5.1	\pm	0.02	$	&	B	&	SIMS	\\
	$	26	$	&	95335-01-01-05	&	$	55326.175	$	&	2010	&	$	0.228	\pm	0.001	$	&	$	7.1	\pm	0.2	$	&	$	4.9	\pm	0.03	$	&	B	&	SIMS	\\
	$	27	$	&	95335-01-01-06	&	$	55326.280	$	&	2010	&	$	0.223	\pm	0.001	$	&	$	6.3	\pm	0.2	$	&	$	4.9	\pm	0.03	$	&	B	&	SIMS	\\
	$	28	$	&	95409-01-18-04	&	$	55327.041	$	&	2010	&	$	0.235	\pm	0.002	$	&	$	8.5	\pm	0.4	$	&	$	4.8	\pm	0.1	$	&	B	&	SIMS	\\
	$	29	$	&	95409-01-18-05	&	$	55327.262	$	&	2010	&	$	0.234	\pm	0.001	$	&	$	7.8	\pm	0.3	$	&	$	4.9	\pm	0.04	$	&	B	&	SIMS	\\
	$	30	$	&	95409-01-19-00	&	$	55330.300	$	&	2010	&	$	0.223	\pm	0.001	$	&	$	6.5	\pm	0.6	$	&	$	4.7	\pm	0.05	$	&	B	&	SIMS	\\
	$	31	$	&	96409-01-04-04	&	$	55585.947	$	&	2010	&	$	0.282	\pm	0.003	$	&	$	10.0	\pm	0.6	$	&	$	2.0	\pm	0.1	$	&	B	&	SIMS	\\
	$	32	$	&	96409-01-04-05	&	$	55586.896	$	&	2010	&	$	0.269	\pm	0.004	$	&	$	9	\pm	2	$	&	$	0.9	\pm	0.1	$	&	B	&	SIMS	\\
	$	33	$	&	96409-01-05-01	&	$	55591.615	$	&	2010	&	$	0.292	\pm	0.003	$	&	$	10	\pm	1	$	&	$	1.7	\pm	0.1	$	&	B	&	SIMS	\\
	$	34	$	&	96409-01-05-02	&	$	55593.502	$	&	2010	&	$	0.299	\pm	0.003	$	&	$	8	\pm	3	$	&	$	1.8	\pm	0.1	$	&	B	&	SIMS	\\
																																		
\hline																																		
\hline																																		
																																		
	$	35	$	&	70109-01-07-00a	&	$	52411.601	$	&	2002	&	$	0.213	\pm	0.001	$	&	$	2.6	\pm	0.1	$	&	$	7.0	\pm	0.5	$	&	A	&	SIMS	\\
	$	36	$	&	70108-03-02-00a	&	$	52419.432	$	&	2002	&	$	0.214	\pm	0.001	$	&	$	2.8	\pm	0.2	$	&	$	6.7	\pm	0.5	$	&	A	&	SIMS	\\
	$	37	$	&	70108-03-02-00a2	&	$	52419.432	$	&	2002	&	$	0.210	\pm	0.001	$	&	$	2.8	\pm	0.2	$	&	$	6.8	\pm	0.4	$	&	A	&	SIMS	\\
	$	38	$	&	70110-01-45-00	&	$	52524.948	$	&	2002	&	$	0.222	\pm	0.001	$	&	$	2.2	\pm	0.2	$	&	$	7.2	\pm	0.6	$	&	A	&	SIMS	\\
	$	39	$	&	70109-01-23-00	&	$	52529.580	$	&	2002	&	$	0.209	\pm	0.001	$	&	$	2.9	\pm	0.2	$	&	$	7.4	\pm	1.3	$	&	A	&	SIMS	\\
	$	40	$	&	70109-01-24-00	&	$	52536.358	$	&	2002	&	$	0.204	\pm	0.001	$	&	$	2.4	\pm	0.2	$	&	$	8.0	\pm	0.9	$	&	A	&	SIMS	\\
	$	41	$	&	92085-01-02-06	&	$	54160.896	$	&	2007	&	$	0.217	\pm	0.001	$	&	$	2.6	\pm	0.2	$	&	$	7.8	\pm	0.7	$	&	A	&	SIMS	\\
	$	42	$	&	92085-01-03-04	&	$	54165.527	$	&	2007	&	$	0.210	\pm	0.001	$	&	$	2.7	\pm	0.1	$	&	$	7.7	\pm	0.7	$	&	A	&	SIMS	\\
																																		
\hline																																		
\hline																																		
																																		
	$	43	$	&	40031-03-02-05	&	$	52388.054	$	&	2002	&	$	0.766	\pm	0.003	$	&	$	29.8	\pm	0.3	$	&	$	0.20	\pm	0.01	$	&	C	&	LHS	\\
	$	44	$	&	70109-01-05-01G	&	$	52391.318	$	&	2002	&	$	0.763	\pm	0.003	$	&	$	29.7	\pm	0.2	$	&	$	0.22	\pm	0.02	$	&	C	&	HIMS	\\
	$	45	$	&	70109-01-06-00	&	$	52400.83	$	&	2002	&	$	0.697	\pm	0.002	$	&	$	22.2	\pm	0.1	$	&	$	1.26	\pm	0.01	$	&	C	&	HIMS	\\
	$	46	$	&	70108-03-01-00	&	$	52400.853	$	&	2002	&	$	0.694	\pm	0.002	$	&	$	22.0	\pm	0.1	$	&	$	1.30	\pm	0.01	$	&	C	&	HIMS	\\
	$	47	$	&	70110-01-10-00	&	$	52402.492	$	&	2002	&	$	0.562	\pm	0.002	$	&	$	20.1	\pm	0.1	$	&	$	4.20	\pm	0.08	$	&	C	&	HIMS	\\
	$	48	$	&	70109-04-01-00	&	$	52405.58	$	&	2002	&	$	0.354	\pm	0.001	$	&	$	15.45	\pm	0.04	$	&	$	5.46	\pm	0.01	$	&	C	&	HIMS	\\
	$	49	$	&	70109-04-01-01	&	$	52405.713	$	&	2002	&	$	0.356	\pm	0.001	$	&	$	15.44	\pm	0.02	$	&	$	5.45	\pm	0.01	$	&	C	&	HIMS	\\
	$	50	$	&	70109-04-01-02	&	$	52406.07	$	&	2002	&	$	0.360	\pm	0.001	$	&	$	15.6	\pm	0.1	$	&	$	5.34	\pm	0.02	$	&	C	&	HIMS	\\
	$	51	$	&	70110-01-11-00	&	$	52406.701	$	&	2002	&	$	0.342	\pm	0.001	$	&	$	15.0	\pm	0.1	$	&	$	5.82	\pm	0.02	$	&	C	&	HIMS	\\
	$	52	$	&	70110-01-12-00	&	$	52410.528	$	&	2002	&	$	0.266	\pm	0.001	$	&	$	11.5	\pm	0.1	$	&	$	8.1	\pm	0.2	$	&	C	&	HIMS	\\
	$	53	$	&	70109-01-37-00	&	$	52694.922	$	&	2002	&	$	0.237	\pm	0.002	$	&	$	10.7	\pm	0.4	$	&	$	8.6	\pm	0.2	$	&	C	&	HIMS	\\
	$	54	$	&	70128-02-02-00	&	$	52696.355	$	&	2002	&	$	0.272	\pm	0.001	$	&	$	12.1	\pm	0.1	$	&	$	8.02	\pm	0.04	$	&	C	&	HIMS	\\
	$	55	$	&	50117-01-03-01	&	$	52706.767	$	&	2002	&	$	0.364	\pm	0.003	$	&	$	19.2	\pm	0.3	$	&	$	6.7	\pm	0.1	$	&	C	&	HIMS	\\
	$	56	$	&	50117-01-03-00	&	$	52706.84	$	&	2002	&	$	0.363	\pm	0.002	$	&	$	18.7	\pm	0.2	$	&	$	6.77	\pm	0.02	$	&	C	&	HIMS	\\
	$	57	$	&	70109-02-01-00	&	$	52709.859	$	&	2002	&	$	0.293	\pm	0.002	$	&	$	14.2	\pm	0.2	$	&	$	8.0	\pm	0.1	$	&	C	&	HIMS	\\
	$	58	$	&	70109-02-01-01	&	$	52709.991	$	&	2002	&	$	0.289	\pm	0.002	$	&	$	14.3	\pm	0.4	$	&	$	8.1	\pm	0.1	$	&	C	&	HIMS	\\
	$	59	$	&	60705-01-56-00	&	$	52710.715	$	&	2002	&	$	0.306	\pm	0.003	$	&	$	15.1	\pm	0.4	$	&	$	7.8	\pm	0.1	$	&	C	&	HIMS	\\
	$	60	$	&	70110-01-94-00	&	$	52724.226	$	&	2002	&	$	0.395	\pm	0.004	$	&	$	18.2	\pm	0.4	$	&	$	6.1	\pm	0.1	$	&	C	&	HIMS	\\
	$	61	$	&	70110-01-95-00	&	$	52727.252	$	&	2002	&	$	0.480	\pm	0.005	$	&	$	22.8	\pm	0.5	$	&	$	4.7	\pm	0.1	$	&	C	&	HIMS	\\
	$	62	$	&	60705-01-59-00	&	$	52731.562	$	&	2002	&	$	0.586	\pm	0.003	$	&	$	25.2	\pm	0.2	$	&	$	2.9	\pm	0.1	$	&	C	&	HIMS	\\
	$	63	$	&	60705-01-68-00	&	$	53218.11	$	&	2004	&	$	0.763	\pm	0.005	$	&	$	30.8	\pm	0.3	$	&	$	0.5	\pm	0.0	$	&	C	&	LHS	\\
	$	64	$	&	60705-01-68-01	&	$	53222.24	$	&	2004	&	$	0.728	\pm	0.004	$	&	$	26.9	\pm	0.2	$	&	$	1.03	\pm	0.04	$	&	C	&	HIMS	\\
	$	65	$	&	60705-01-69-00	&	$	53225.40	$	&	2004	&	$	0.714	\pm	0.003	$	&	$	24.9	\pm	0.2	$	&	$	1.3	\pm	0.0	$	&	C	&	HIMS	\\
	$	66	$	&	90704-01-01-00	&	$	53226.43	$	&	2004	&	$	0.705	\pm	0.003	$	&	$	24.9	\pm	0.1	$	&	$	2.0	\pm	0.1	$	&	C	&	HIMS	\\
	$	67	$	&	60705-01-69-01	&	$	53228.99	$	&	2004	&	$	0.646	\pm	0.003	$	&	$	23.7	\pm	0.2	$	&	$	2.9	\pm	0.2	$	&	C	&	HIMS	\\
	$	68	$	&	60705-01-70-00	&	$	53230.96	$	&	2004	&	$	0.437	\pm	0.002	$	&	$	18.3	\pm	0.2	$	&	$	4.3	\pm	0.1	$	&	C	&	HIMS	\\
	$	69	$	&	90110-02-01-02	&	$	53232.34	$	&	2004	&	$	0.386	\pm	0.002	$	&	$	17.5	\pm	0.1	$	&	$	5.2	\pm	0.1	$	&	C	&	HIMS	\\
	$	70	$	&	90110-02-01-00	&	$	53232.40	$	&	2004	&	$	0.361	\pm	0.001	$	&	$	16.35	\pm	0.04	$	&	$	5.8	\pm	0.1	$	&	C	&	HIMS	\\
	$	71	$	&	90704-01-11-00	&	$	53472.33	$	&	2004	&	$	0.584	\pm	0.004	$	&	$	24.9	\pm	0.3	$	&	$	2.7	\pm	0.1	$	&	C	&	HIMS	\\
	$	72	$	&	92035-01-02-01	&	$	54133.922	$	&	2007	&	$	0.771	\pm	0.003	$	&	$	30.8	\pm	0.2	$	&	$	0.28	\pm	0.01	$	&	C	&	LHS	\\
	$	73	$	&	92035-01-02-02	&	$	54135.033	$	&	2007	&	$	0.771	\pm	0.003	$	&	$	30.8	\pm	0.2	$	&	$	0.30	\pm	0.01	$	&	C	&	LHS	\\
	$	74	$	&	92035-01-02-03	&	$	54136.015	$	&	2007	&	$	0.766	\pm	0.003	$	&	$	30.1	\pm	0.2	$	&	$	0.37	\pm	0.01	$	&	C	&	LHS	\\
	$	75	$	&	92035-01-02-04	&	$	54136.997	$	&	2007	&	$	0.759	\pm	0.003	$	&	$	29.8	\pm	0.2	$	&	$	0.43	\pm	0.01	$	&	C	&	LHS	\\
	$	76	$	&	92035-01-02-08	&	$	54137.851	$	&	2007	&	$	0.748	\pm	0.003	$	&	$	27.5	\pm	0.3	$	&	$	0.55	\pm	0.02	$	&	C	&	HIMS	\\
	$	77	$	&	92035-01-02-07	&	$	54138.83	$	&	2007	&	$	0.731	\pm	0.002	$	&	$	25.8	\pm	0.2	$	&	$	0.90	\pm	0.01	$	&	C	&	HIMS	\\
	$	78	$	&	92035-01-02-06	&	$	54139.942	$	&	2007	&	$	0.686	\pm	0.002	$	&	$	22.4	\pm	0.1	$	&	$	0.99	\pm	0.01	$	&	C	&	HIMS	\\
	$	79	$	&	92035-01-03-00	&	$	54140.204	$	&	2007	&	$	0.670	\pm	0.002	$	&	$	21.7	\pm	0.1	$	&	$	1.13	\pm	0.01	$	&	C	&	HIMS	\\
	$	80	$	&	92035-01-03-01	&	$	54141.055	$	&	2007	&	$	0.621	\pm	0.002	$	&	$	20.7	\pm	0.1	$	&	$	1.68	\pm	0.01	$	&	C	&	HIMS	\\
	$	81	$	&	92035-01-03-02	&	$	54142.036	$	&	2007	&	$	0.547	\pm	0.002	$	&	$	19.6	\pm	0.1	$	&	$	2.45	\pm	0.01	$	&	C	&	HIMS	\\
	$	82	$	&	92035-01-03-03	&	$	54143.019	$	&	2007	&	$	0.461	\pm	0.002	$	&	$	18.3	\pm	0.1	$	&	$	3.52	\pm	0.01	$	&	C	&	HIMS	\\
	$	83	$	&	92428-01-04-00	&	$	54143.870	$	&	2007	&	$	0.411	\pm	0.001	$	&	$	17.1	\pm	0.1	$	&	$	4.34	\pm	0.02	$	&	C	&	HIMS	\\
	$	84	$	&	92428-01-04-01	&	$	54143.951	$	&	2007	&	$	0.419	\pm	0.001	$	&	$	17.2	\pm	0.1	$	&	$	4.23	\pm	0.02	$	&	C	&	HIMS	\\
	$	85	$	&	92428-01-04-02	&	$	54144.086	$	&	2007	&	$	0.424	\pm	0.002	$	&	$	17.4	\pm	0.1	$	&	$	4.13	\pm	0.03	$	&	C	&	HIMS	\\
	$	86	$	&	92428-01-04-03	&	$	54144.871	$	&	2007	&	$	0.380	\pm	0.001	$	&	$	16.4	\pm	0.1	$	&	$	4.99	\pm	0.03	$	&	C	&	HIMS	\\
	$	87	$	&	92035-01-03-05	&	$	54145.114	$	&	2007	&	$	0.343	\pm	0.001	$	&	$	14.9	\pm	0.1	$	&	$	5.80	\pm	0.03	$	&	C	&	HIMS	\\
	$	88	$	&	92085-01-03-00	&	$	54161.669	$	&	2007	&	$	0.295	\pm	0.001	$	&	$	12.9	\pm	0.1	$	&	$	7.1	\pm	0.1	$	&	C	&	HIMS	\\
	$	89	$	&	92085-01-03-02	&	$	54163.698	$	&	2007	&	$	0.288	\pm	0.001	$	&	$	12.5	\pm	0.1	$	&	$	7.3	\pm	0.2	$	&	C	&	HIMS	\\
	$	90	$	&	92085-01-03-03	&	$	54164.557	$	&	2007	&	$	0.296	\pm	0.001	$	&	$	12.4	\pm	0.1	$	&	$	7.0	\pm	0.2	$	&	C	&	HIMS	\\
	$	91	$	&	92704-03-11-00	&	$	54234.839	$	&	2007	&	$	0.493	\pm	0.004	$	&	$	21.3	\pm	0.4	$	&	$	4.0	\pm	0.1	$	&	C	&	HIMS	\\
	$	92	$	&	92704-03-11-01	&	$	54235.791	$	&	2007	&	$	0.549	\pm	0.005	$	&	$	22.1	\pm	0.4	$	&	$	3.3	\pm	0.1	$	&	C	&	HIMS	\\
	$	93	$	&	92704-04-01-01	&	$	54236.446	$	&	2007	&	$	0.577	\pm	0.004	$	&	$	22.8	\pm	0.6	$	&	$	3.0	\pm	0.2	$	&	C	&	HIMS	\\
	$	94	$	&	92704-04-01-02	&	$	54236.513	$	&	2007	&	$	0.587	\pm	0.004	$	&	$	24.8	\pm	0.4	$	&	$	2.7	\pm	0.1	$	&	C	&	HIMS	\\
	$	95	$	&	92704-03-12-00	&	$	54236.591	$	&	2007	&	$	0.598	\pm	0.005	$	&	$	25.5	\pm	1.1	$	&	$	2.7	\pm	0.4	$	&	C	&	HIMS	\\
	$	96	$	&	92704-04-01-04	&	$	54237.356	$	&	2007	&	$	0.593	\pm	0.004	$	&	$	25.2	\pm	0.5	$	&	$	2.8	\pm	0.1	$	&	C	&	HIMS	\\
	$	97	$	&	92704-04-01-05	&	$	54237.421	$	&	2007	&	$	0.584	\pm	0.004	$	&	$	24.5	\pm	0.4	$	&	$	3.1	\pm	0.3	$	&	C	&	HIMS	\\
	$	98	$	&	92704-03-12-01	&	$	54237.488	$	&	2007	&	$	0.596	\pm	0.004	$	&	$	24.8	\pm	0.3	$	&	$	2.7	\pm	0.1	$	&	C	&	HIMS	\\
	$	99	$	&	95409-01-12-04	&	$	55286.727	$	&	2010	&	$	0.787	\pm	0.003	$	&	$	32.4	\pm	0.4	$	&	$	0.22	\pm	0.01	$	&	C	&	LHS	\\
	$	100	$	&	95409-01-13-03	&	$	55288.367	$	&	2010	&	$	0.783	\pm	0.003	$	&	$	31.5	\pm	0.3	$	&	$	0.2	\pm	0.1	$	&	C	&	LHS	\\
	$	101	$	&	95409-01-13-00	&	$	55289.618	$	&	2010	&	$	0.777	\pm	0.003	$	&	$	31.0	\pm	0.3	$	&	$	0.26	\pm	0.01	$	&	C	&	LHS	\\
	$	102	$	&	95409-01-13-04	&	$	55290.722	$	&	2010	&	$	0.781	\pm	0.003	$	&	$	31.7	\pm	0.2	$	&	$	0.29	\pm	0.01	$	&	C	&	LHS	\\
	$	103	$	&	95409-01-13-02	&	$	55291.649	$	&	2010	&	$	0.775	\pm	0.003	$	&	$	31.6	\pm	0.3	$	&	$	0.32	\pm	0.01	$	&	C	&	LHS	\\
	$	104	$	&	95409-01-13-05	&	$	55292.779	$	&	2010	&	$	0.777	\pm	0.003	$	&	$	30.9	\pm	0.4	$	&	$	0.38	\pm	0.02	$	&	C	&	LHS	\\
	$	105	$	&	95409-01-13-01	&	$	55293.088	$	&	2010	&	$	0.772	\pm	0.003	$	&	$	30.6	\pm	0.4	$	&	$	0.38	\pm	0.05	$	&	C	&	LHS	\\
	$	106	$	&	95409-01-13-06	&	$	55294.124	$	&	2010	&	$	0.770	\pm	0.003	$	&	$	30.1	\pm	0.4	$	&	$	0.47	\pm	0.02	$	&	C	&	LHS	\\
	$	107	$	&	95409-01-14-01	&	$	55296.248	$	&	2010	&	$	0.734	\pm	0.003	$	&	$	26.7	\pm	0.3	$	&	$	1.04	\pm	0.01	$	&	C	&	HIMS	\\
	$	108	$	&	95409-01-14-02	&	$	55297.87	$	&	2010	&	$	0.672	\pm	0.002	$	&	$	21.9	\pm	0.1	$	&	$	1.25	\pm	0.01	$	&	C	&	HIMS	\\
	$	109	$	&	95409-01-14-03	&	$	55298.70	$	&	2010	&	$	0.648	\pm	0.003	$	&	$	21.6	\pm	0.2	$	&	$	1.59	\pm	0.01	$	&	C	&	HIMS	\\
	$	110	$	&	95409-01-14-06	&	$	55299.766	$	&	2010	&	$	0.564	\pm	0.002	$	&	$	20.0	\pm	0.1	$	&	$	2.43	\pm	0.01	$	&	C	&	HIMS	\\
	$	111	$	&	95409-01-14-04	&	$	55300.336	$	&	2010	&	$	0.563	\pm	0.002	$	&	$	20.0	\pm	0.2	$	&	$	2.38	\pm	0.01	$	&	C	&	HIMS	\\
	$	112	$	&	95409-01-14-07	&	$	55300.923	$	&	2010	&	$	0.515	\pm	0.002	$	&	$	19.4	\pm	0.1	$	&	$	2.92	\pm	0.0	$	&	C	&	HIMS	\\
	$	113	$	&	95409-01-14-05	&	$	55301.789	$	&	2010	&	$	0.454	\pm	0.002	$	&	$	17.7	\pm	0.2	$	&	$	3.64	\pm	0.02	$	&	C	&	HIMS	\\
	$	114	$	&	95409-01-15-00	&	$	55302.196	$	&	2010	&	$	0.425	\pm	0.002	$	&	$	17.5	\pm	0.2	$	&	$	4.15	\pm	0.03	$	&	C	&	HIMS	\\
	$	115	$	&	95409-01-15-01	&	$	55303.604	$	&	2010	&	$	0.346	\pm	0.001	$	&	$	14.6	\pm	0.1	$	&	$	5.65	\pm	0.04	$	&	C	&	HIMS	\\
	$	116	$	&	95409-01-17-02	&	$	55318.441	$	&	2010	&	$	0.308	\pm	0.001	$	&	$	13.7	\pm	0.1	$	&	$	6.67	\pm	0.19	$	&	C	&	HIMS	\\
	$	117	$	&	96409-01-06-01	&	$	55598.700	$	&	2010	&	$	0.491	\pm	0.004	$	&	$	26.7	\pm	1.5	$	&	$	4.52	\pm	0.30	$	&	C	&	HIMS	\\																																			
\end{longtable}
\end{center} 

\twocolumn

\onecolumn
\begin{center} 
\begin{longtable}{|c|c|c|c|c|c|c|c|c|c|} 
\caption{Columns are: observation number, reduced $\chi ^2 $,inner disc temperature (kT), inner disc radius R (assuming a distance of 10 kpc and an inclination of 30 $^o$), photon index $\Gamma$, fold Energy E$_{fold}$ (corresponding to high energy cutoff), total flux, hard flux and disk flux calculated in the 2 - 20 keV band and expressed in units of Crab flux.}\label{tab:parameters} \\ 
\endfirsthead

\multicolumn{10}{c}%
{{\tablename\ \thetable{} -- continued from previous page}} \\ \hline
$\#$	&	reduced $\chi ^2 $	&		T$_{Inn radius}$					&		R$^{Inn}_{Disk}$					&		$\Gamma$					&		E$_{fold}$						&	F$_{tot}$/F$_{Crab}$	&	F$_{hard}$/F$_{Crab}$	&	F$_{disk}$/F$_{Crab}$	&	instruments	\\
\hline																																
\hline	
\endhead

\hline \multicolumn{10}{c}{{Continued on next page}} \\
\endfoot

\hline 
\endlastfoot
\hline		

\hline																																									
$\#$	&	reduced $\chi ^2 $	&		T$_{Inn radius}$					&		R$^{Inn}_{Disk}$					&		$\Gamma$					&		E$_{fold}$						&	F$_{tot}$/F$_{Crab}$	&	F$_{hard}$/F$_{Crab}$	&	F$_{disk}$/F$_{Crab}$	&	instruments	\\	
	\hline																																									
\hline																																									
																																								
1	&	0.92	&	$	0.94	_{-	0.04	}^{+	0.02	}$ &	$	43	_{-	2	}^{+	3	}$ &	$	2.6	_{-	0.1	}^{+	0.1	}$ &		 -						&	0.501		&	0.063		&	0.321		&	PCA+HEXTE	\\
2	&	1.01	&	$	0.93	_{-	0.04	}^{+	0.04	}$ &	$	41	_{-	4	}^{+	5	}$ &	$	2.6	_{-	0.1	}^{+	0.2	}$ &		 -						&	0.462		&	0.070		&	0.270		&	PCA+HEXTE	\\
3	&	1.11	&	$	0.93	_{-	0.04	}^{+	0.05	}$ &	$	38	_{-	4	}^{+	5	}$ &	$	2.8	_{-	0.2	}^{+	0.2	}$ &		 -						&	0.434		&	0.066		&	0.235		&	PCA+HEXTE	\\
4	&	0.81	&	$	0.90	_{-	0.03	}^{+	0.04	}$ &	$	43	_{-	4	}^{+	6	}$ &	$	2.6	_{-	0.1	}^{+	0.2	}$ &		 -						&	0.395		&	0.048		&	0.259		&	PCA+HEXTE	\\
5	&	0.96	&	$	0.91	_{-	0.03	}^{+	0.04	}$ &	$	45	_{-	4	}^{+	4	}$ &	$	3.0	_{-	0.2	}^{+	0.2	}$ &		 -						&	0.494		&	0.060		&	0.297		&	PCA+HEXTE	\\
6	&	0.81	&	$	0.61	_{-	0.15	}^{+	0.06	}$ &	$	33	_{-	7	}^{+	31	}$ &	$	3.8	_{-	0.3	}^{+	0.2	}$ &		 -						&	0.027		&	0.004		&	0.016		&	PCA+HEXTE	\\
7	&	0.85	&	$	0.80	_{-	0.03	}^{+	0.06	}$ &	$	37	_{-	6	}^{+	4	}$ &	$	2.7	_{-	0.4	}^{+	0.2	}$ &		 -						&	0.178		&	0.027		&	0.104		&	PCA+HEXTE	\\
8	&	0.84	&	$	0.85	_{-	0.03	}^{+	0.04	}$ &	$	31	_{-	3	}^{+	3	}$ &	$	2.6	_{-	0.1	}^{+	0.1	}$ &		 -						&	0.175		&	0.029		&	0.099		&	PCA+HEXTE	\\
9	&	1.07	&	$	0.90	_{-	0.03	}^{+	0.05	}$ &	$	38	_{-	4	}^{+	3	}$ &	$	2.5	_{-	0.1	}^{+	0.1	}$ &		 -						&	0.314		&	0.047		&	0.195		&	PCA+HEXTE	\\
10	&	0.93	&	$	0.81	_{-	0.07	}^{+	0.11	}$ &	$	19	_{-	4	}^{+	5	}$ &	$	2.4	_{-	0.4	}^{+	0.4	}$ &		 -						&	0.056		&	0.011		&	0.030		&	PCA+HEXTE	\\
11	&	0.92	&	$	0.93	_{-	0.02	}^{+	0.04	}$ &	$	46	_{-	4	}^{+	3	}$ &	$	2.7	_{-	0.03	}^{+	0.1	}$ &		 -						&	0.653		&	0.099		&	0.367		&	PCA+HEXTE	\\
12	&	1.08	&	$	0.92	_{-	0.03	}^{+	0.03	}$ &	$	44	_{-	3	}^{+	4	}$ &	$	2.7	_{-	0.1	}^{+	0.1	}$ &		 -						&	0.536		&	0.080		&	0.307		&	PCA+HEXTE	\\
13	&	0.89	&	$	0.76	_{-	0.04	}^{+	0.05	}$ &	$	25	_{-	4	}^{+	4	}$ &	$	2.2	_{-	0.3	}^{+	0.3	}$ &		 -						&	0.054		&	0.008		&	0.035		&	PCA+HEXTE	\\
14	&	0.99	&	$	0.78	_{-	0.05	}^{+	0.04	}$ &	$	22	_{-	3	}^{+	5	}$ &	$	2.1	_{-	0.2	}^{+	0.3	}$ &		 -						&	0.053		&	0.009		&	0.033		&	PCA+HEXTE	\\
15	&	1.04	&	$	0.8	_{-	0.1	}^{+	0.1	}$ &	$	16	_{-	3	}^{+	3	}$ &	$	2.1	_{-	0.2	}^{+	0.3	}$ &		 -						&	0.046		&	0.010		&	0.025		&	PCA+HEXTE	\\
16	&	1.26	&	$	0.96	_{-	0.04	}^{+	0.03	}$ &	$	40	_{-	3	}^{+	4	}$ &	$	2.4	_{-	0.04	}^{+	0.04	}$ &		 -						&	0.496		&	0.071		&	0.318		&	PCA+HEXTE *	\\
17	&	1.18	&	$	0.90	_{-	0.03	}^{+	0.02	}$ &	$	48	_{-	3	}^{+	4	}$ &	$	2.5	_{-	0.05	}^{+	0.1	}$ &		 -						&	0.492		&	0.062		&	0.329		&	PCA+HEXTE *	\\
18	&	1.00	&	$	0.94	_{-	0.04	}^{+	0.03	}$ &	$	38	_{-	3	}^{+	4	}$ &	$	2.5	_{-	0.1	}^{+	0.1	}$ &		 -						&	0.438		&	0.068		&	0.261		&	PCA+HEXTE *	\\
19	&	1.25	&	$	0.91	_{-	0.02	}^{+	0.04	}$ &	$	43	_{-	4	}^{+	3	}$ &	$	2.3	_{-	0.04	}^{+	0.1	}$ &		 -						&	0.432		&	0.064		&	0.275		&	PCA+HEXTE *	\\
20	&	1.23	&	$	0.93	_{-	0.03	}^{+	0.04	}$ &	$	39	_{-	4	}^{+	3	}$ &	$	2.4	_{-	0.1	}^{+	0.1	}$ &		 -						&	0.382		&	0.052		&	0.252		&	PCA+HEXTE *	\\
21	&	1.54	&	$	0.89	_{-	0.04	}^{+	0.04	}$ &	$	44	_{-	5	}^{+	4	}$ &	$	2.36	_{-	0.04	}^{+	0.04	}$ &		 -						&	0.383		&	0.049		&	0.263		&	PCA+HEXTE *	\\
22	&	1.74	&	$	0.93	_{-	0.03	}^{+	0.04	}$ &	$	38	_{-	4	}^{+	4	}$ &	$	2.38	_{-	0.04	}^{+	0.04	}$ &		 -						&	0.384		&	0.057		&	0.242		&	PCA+HEXTE *	\\
23	&	1.07	&	$	0.89	_{-	0.03	}^{+	0.03	}$ &	$	41	_{-	4	}^{+	5	}$ &	$	2.7	_{-	0.4	}^{+	0.4	}$ &		 -						&	0.391		&	0.056		&	0.225		&	PCA+HEXTE *	\\
24	&	1.07	&	$	0.87	_{-	0.01	}^{+	0.04	}$ &	$	43	_{-	4	}^{+	4	}$ &	$	2.7	_{-	0.2	}^{+	0.2	}$ &		 -						&	0.385		&	0.054		&	0.226		&	PCA+HEXTE *	\\
25	&	1.12	&	$	0.86	_{-	0.03	}^{+	0.04	}$ &	$	45	_{-	4	}^{+	3	}$ &	$	2.7	_{-	0.2	}^{+	0.2	}$ &		 -						&	0.380		&	0.051		&	0.231		&	PCA+HEXTE *	\\
26	&	1.03	&	$	0.84	_{-	0.03	}^{+	0.04	}$ &	$	47	_{-	5	}^{+	5	}$ &	$	2.9	_{-	0.4	}^{+	0.2	}$ &		 -						&	0.371		&	0.046		&	0.219		&	PCA+HEXTE *	\\
27	&	1.28	&	$	0.85	_{-	0.03	}^{+	0.04	}$ &	$	46	_{-	5	}^{+	5	}$ &	$	2.7	_{-	0.3	}^{+	0.3	}$ &		 -						&	0.357		&	0.043		&	0.230		&	PCA+HEXTE *	\\
28	&	1.26	&	$	0.89	_{-	0.04	}^{+	0.05	}$ &	$	41	_{-	5	}^{+	5	}$ &	$	2.3	_{-	0.1	}^{+	0.1	}$ &		 -						&	0.327		&	0.042		&	0.230		&	PCA+HEXTE *	\\
29	&	1.50	&	$	0.90	_{-	0.05	}^{+	0.05	}$ &	$	40	_{-	4	}^{+	7	}$ &	$	2.4	_{-	0.03	}^{+	0.1	}$ &		 -						&	0.342		&	0.045		&	0.232		&	PCA+HEXTE *	\\
30	&	1.21	&	$	0.86	_{-	0.03	}^{+	0.04	}$ &	$	46	_{-	4	}^{+	5	}$ &	$	2.4	_{-	0.1	}^{+	0.1	}$ &		 -						&	0.330		&	0.039		&	0.232		&	PCA+HEXTE *	\\
31	&	1.27	&	$	0.71	_{-	0.06	}^{+	0.06	}$ &	$	31	_{-	6	}^{+	14	}$ &	$	2.0	_{-	0.1	}^{+	0.2	}$ &		 -						&	0.055		&	0.009		&	0.038		&	PCA+HEXTE *	\\
32	&	1.78	&	$	0.94	_{-	0.03	}^{+	0.14	}$ &	$	13	_{-	3	}^{+	2	}$ &	$	1.4	_{-	0.1	}^{+	0.1	}$ &		 -						&	0.039		&	0.006		&	0.029		&	PCA+HEXTE *	\\
33	&	0.94	&	$	0.77	_{-	0.07	}^{+	0.06	}$ &	$	22	_{-	4	}^{+	4	}$ &	$	1.8	_{-	0.1	}^{+	0.1	}$ &		 -						&	0.042		&	0.007		&	0.028		&	PCA+HEXTE *	\\
34	&	1.31	&	$	0.85	_{-	0.04	}^{+	0.06	}$ &	$	15	_{-	2	}^{+	2	}$ &	$	1.8	_{-	0.1	}^{+	0.1	}$ &		 -						&	0.035		&	0.007		&	0.023		&	PCA+HEXTE *	\\

35	&	1.09	&	$	0.94	_{-	0.04	}^{+	0.02	}$ &	$	44	_{-	2	}^{+	6	}$ &	$	2.6	_{-	0.1	}^{+	0.1	}$ &		 -						&	0.488		&	0.055		&	0.336		&	PCA+HEXTE	\\
36	&	1.31	&	$	0.90	_{-	0.01	}^{+	0.02	}$ &	$	44	_{-	3	}^{+	3	}$ &	$	2.6	_{-	0.1	}^{+	0.1	}$ &		 -						&	0.394		&	0.045		&	0.271		&	PCA+HEXTE	\\
37	&	1.31	&	$	0.90	_{-	0.02	}^{+	0.02	}$ &	$	44	_{-	3	}^{+	3	}$ &	$	2.6	_{-	0.1	}^{+	0.1	}$ &		 -						&	0.394		&	0.045		&	0.271		&	PCA+HEXTE	\\
38	&	0.97	&	$	0.95	_{-	0.04	}^{+	0.04	}$ &	$	44	_{-	4	}^{+	5	}$ &	$	2.7	_{-	0.2	}^{+	0.1	}$ &		 -						&	0.531		&	0.065		&	0.342		&	PCA+HEXTE	\\
39	&	0.85	&	$	0.95	_{-	0.03	}^{+	0.02	}$ &	$	43	_{-	2	}^{+	3	}$ &	$	2.6	_{-	0.1	}^{+	0.1	}$ &		 -						&	0.480		&	0.053		&	0.335		&	PCA+HEXTE	\\
40	&	0.89	&	$	0.90	_{-	0.02	}^{+	0.03	}$ &	$	48	_{-	4	}^{+	3	}$ &	$	2.6	_{-	0.1	}^{+	0.1	}$ &		 -						&	0.448		&	0.047		&	0.311		&	PCA+HEXTE	\\
41	&	0.86	&	$	0.90	_{-	0.02	}^{+	0.04	}$ &	$	51	_{-	4	}^{+	2	}$ &	$	2.7	_{-	0.1	}^{+	0.1	}$ &		 -						&	0.535		&	0.060		&	0.357		&	PCA+HEXTE	\\
42	&	1.08	&	$	0.88	_{-	0.02	}^{+	0.04	}$ &	$	49	_{-	4	}^{+	3	}$ &	$	2.8	_{-	0.2	}^{+	0.1	}$ &		 -						&	0.461		&	0.050		&	0.306		&	PCA+HEXTE	\\

43	&	1.19	&			 -				&			 -				&	$	1.74	_{-	0.02	}^{+	0.02	}$ &	$	119	_{-	7	}^{+	8	}$	&	0.401		&	0.236		&	0.000		&	PCA+HEXTE	\\
44	&	0.88	&			 -				&			 -				&	$	1.78	_{-	0.02	}^{+	0.02	}$ &	$	133	_{-	9	}^{+	10	}$	&	0.407		&	0.238		&	0.000		&	PCA+HEXTE	\\
45	&	0.91	&			 -				&			 -				&	$	1.96	_{-	0.02	}^{+	0.02	}$ &	$	122	_{-	10	}^{+	11	}$	&	0.430		&	0.230		&	0.000		&	PCA+HEXTE	\\
46	&	1.50	&			 -				&			 -				&	$	1.98	_{-	0.02	}^{+	0.02	}$ &	$	129	_{-	8	}^{+	9	}$	&	0.428		&	0.227		&	0.000		&	PCA+HEXTE	\\
47	&	0.84	&			 -				&			 -				&	$	2.45	_{-	0.03	}^{+	0.03	}$ &			 -					&	0.410		&	0.173		&	0.000		&	PCA+HEXTE	\\
48	&	0.80	&	$	1.01	_{-	0.05	}^{+	0.05	}$ &	$	26	_{-	2	}^{+	3	}$ &	$	2.59	_{-	0.22	}^{+	0.07	}$ &			 -					&	0.494		&	0.122		&	0.164		&	PCA+HEXTE	\\
49	&	1.12	&	$	0.99	_{-	0.04	}^{+	0.04	}$ &	$	27	_{-	2	}^{+	3	}$ &	$	2.56	_{-	0.15	}^{+	0.06	}$ &			 -					&	0.499		&	0.123		&	0.171		&	PCA+HEXTE	\\
50	&	0.85	&	$	1.0	_{-	0.1	}^{+	0.1	}$ &	$	24	_{-	4	}^{+	3	}$ &	$	2.7	_{-	0.1	}^{+	0.1	}$ &			 -					&	0.471		&	0.119		&	0.140		&	PCA+HEXTE	\\
51	&	0.84	&	$	1.03	_{-	0.06	}^{+	0.06	}$ &	$	28	_{-	3	}^{+	3	}$ &	$	2.2	_{-	0.2	}^{+	0.4	}$ &			 -					&	0.460		&	0.106		&	0.215		&	PCA+HEXTE	\\
52	&	0.94	&	$	0.94	_{-	0.03	}^{+	0.04	}$ &	$	39	_{-	3	}^{+	4	}$ &	$	2.5	_{-	0.2	}^{+	0.1	}$ &			 -					&	0.467		&	0.078		&	0.259		&	PCA+HEXTE	\\
53	&	0.89	&	$	0.69	_{-	0.05	}^{+	0.05	}$ &	$	30	_{-	5	}^{+	8	}$ &	$	2.4	_{-	0.2	}^{+	0.6	}$ &			 -					&	0.046		&	0.006		&	0.030		&	PCA+HEXTE	\\
54	&	1.36	&	$	0.73	_{-	0.03	}^{+	0.03	}$ &	$	23	_{-	3	}^{+	3	}$ &	$	2.6	_{-	0.1	}^{+	0.1	}$ &			 -					&	0.041		&	0.007		&	0.022		&	PCA+HEXTE	\\
55	&	0.67	&	$	0.9	_{-	0.1	}^{+	0.1	}$ &	$	11	_{-	2	}^{+	6	}$ &	$	2.5	_{-	0.4	}^{+	0.2	}$ &			 -					&	0.034		&	0.009		&	0.011		&	PCA+HEXTE	\\
56	&	1.11	&	$	0.78	_{-	0.06	}^{+	0.07	}$ &	$	15	_{-	3	}^{+	4	}$ &	$	2.5	_{-	0.1	}^{+	0.2	}$ &			 -					&	0.035		&	0.009		&	0.013		&	PCA+HEXTE	\\
57	&	0.97	&	$	0.75	_{-	0.04	}^{+	0.04	}$ &	$	19	_{-	2	}^{+	3	}$ &	$	2.2	_{-	0.2	}^{+	0.2	}$ &			 -					&	0.032		&	0.006		&	0.019		&	PCA+HEXTE	\\
58	&	0.74	&	$	0.7	_{-	0.1	}^{+	0.1	}$ &	$	27	_{-	5	}^{+	8	}$ &	$	2.4	_{-	0.4	}^{+	0.3	}$ &			 -					&	0.035		&	0.006		&	0.021		&	PCA+HEXTE	\\
59	&	1.01	&	$	0.7	_{-	0.1	}^{+	0.1	}$ &	$	21	_{-	6	}^{+	6	}$ &	$	2.8	_{-	0.4	}^{+	0.4	}$ &			 -					&	0.026		&	0.005		&	0.013		&	PCA+HEXTE	\\
60	&	1.29	&			 -				&			 -				&	$	3.2	_{-	0.1	}^{+	0.1	}$ &			 -					&	0.035		&	0.010		&	0.000		&	PCA+HEXTE	\\
61	&	1.03	&			 -				&			 -				&	$	2.9	_{-	0.1	}^{+	0.1	}$ &			 -					&	0.034		&	0.012		&	0.000		&	PCA+HEXTE	\\
62	&	1.17	&			 -				&			 -				&	$	2.48	_{-	0.06	}^{+	0.03	}$ &			 -					&	0.033		&	0.015		&	0.000		&	PCA+HEXTE	\\
63	&	1.02	&			 -				&			 -				&	$	1.83	_{-	0.03	}^{+	0.03	}$ &			 -					&	0.102		&	0.061		&	0.000		&	PCA+HEXTE	\\
64	&	1.15	&			 -				&			 -				&	$	1.98	_{-	0.03	}^{+	0.03	}$ &			 -					&	0.126		&	0.070		&	0.000		&	PCA+HEXTE	\\
65	&	1.20	&			 -				&			 -				&	$	1.97	_{-	0.01	}^{+	0.03	}$ &			 -					&	0.132		&	0.073		&	0.000		&	PCA+HEXTE	\\
66	&	1.40	&			 -				&			 -				&	$	2.01	_{-	0.02	}^{+	0.03	}$ &			 -					&	0.136		&	0.074		&	0.000		&	PCA+HEXTE	\\
67	&	0.90	&			 -				&			 -				&	$	2.23	_{-	0.03	}^{+	0.03	}$ &			 -					&	0.134		&	0.066		&	0.000		&	PCA+HEXTE	\\
68	&	0.94	&	$	1.2	_{-	0.1	}^{+	0.1	}$ &	$	8	_{-	1	}^{+	6	}$ &	$	2.3	_{-	0.1	}^{+	0.1	}$ &			 -					&	0.124		&	0.040		&	0.038		&	PCA+HEXTE	\\
69	&	1.09	&	$	1.17	_{-	0.09	}^{+	0.09	}$ &	$	11	_{-	1	}^{+	2	}$ &	$	1.7	_{-	0.1	}^{+	0.7	}$ &	$	64	_{-	17	}^{+	13	}$	&	0.127		&	0.035		&	0.065		&	PCA+HEXTE	\\
70	&	0.88	&	$	1.13	_{-	0.08	}^{+	0.05	}$ &	$	12	_{-	1	}^{+	1	}$ &	$	1.9	_{-	0.2	}^{+	0.4	}$ &	$	102	_{-	38	}^{+	11	}$	&	0.127		&	0.033		&	0.064		&	PCA+HEXTE	\\
71	&	1.14	&			 -				&			 -				&	$	2.4	_{-	0.1	}^{+	0.1	}$ &			 -					&	0.046		&	0.021		&	0.000		&	PCA+HEXTE	\\
72	&	1.30	&			 -				&			 -				&	$	1.82	_{-	0.02	}^{+	0.02	}$ &	$	163	_{-	14	}^{+	15	}$	&	0.367		&	0.214		&	0.000		&	PCA+HEXTE	\\
73	&	1.19	&			 -				&			 -				&	$	1.80	_{-	0.02	}^{+	0.02	}$ &	$	139	_{-	9	}^{+	11	}$	&	0.371		&	0.216		&	0.000		&	PCA+HEXTE	\\
74	&	1.52	&			 -				&			 -				&	$	1.83	_{-	0.02	}^{+	0.02	}$ &	$	146	_{-	10	}^{+	13	}$	&	0.386		&	0.222		&	0.000		&	PCA+HEXTE	\\
75	&	1.25	&			 -				&			 -				&	$	1.84	_{-	0.02	}^{+	0.02	}$ &	$	137	_{-	9	}^{+	11	}$	&	0.413		&	0.236		&	0.000		&	PCA+HEXTE	\\
76	&	1.17	&			 -				&			 -				&	$	1.86	_{-	0.02	}^{+	0.02	}$ &	$	131	_{-	12	}^{+	16	}$	&	0.424		&	0.239		&	0.000		&	PCA+HEXTE	\\
77	&	1.13	&			 -				&			 -				&	$	1.92	_{-	0.02	}^{+	0.02	}$ &	$	140	_{-	11	}^{+	13	}$	&	0.433		&	0.238		&	0.000		&	PCA+HEXTE	\\
78	&	1.03	&			 -				&			 -				&	$	2.05	_{-	0.02	}^{+	0.02	}$ &	$	142	_{-	11	}^{+	14	}$	&	0.434		&	0.223		&	0.000		&	PCA+HEXTE	\\
79	&	1.05	&			 -				&			 -				&	$	2.09	_{-	0.02	}^{+	0.02	}$ &	$	164	_{-	16	}^{+	19	}$	&	0.436		&	0.219		&	0.000		&	PCA+HEXTE	\\
80	&	0.81	&			 -				&			 -				&	$	2.25	_{-	0.02	}^{+	0.02	}$ &			 -					&	0.443		&	0.206		&	0.000		&	PCA+HEXTE	\\
81	&	0.92	&			 -				&			 -				&	$	2.539	_{-	0.004	}^{+	0.004	}$ &			 -					&	0.475		&	0.191		&	0.000		&	PCA+HEXTE	\\
82	&	1.00	&	$	1.0	_{-	0.1	}^{+	0.0	}$ &	$	19	_{-	3	}^{+	6	}$ &	$	2.46	_{-	0.22	}^{+	0.14	}$ &			 -					&	0.501		&	0.166		&	0.085		&	PCA+HEXTE	\\
83	&	1.00	&	$	0.980	_{-	0.004	}^{+	0.0	}$ &	$	23.4	_{-	0.2	}^{+	0.2	}$ &	$	2.589	_{-	0.003	}^{+	0.003	}$ &			 -					&	0.533		&	0.154		&	0.121		&	PCA+HEXTE	\\
84	&	0.96	&	$	1.035	_{-	0.005	}^{+	0.0	}$ &	$	18.8	_{-	0.2	}^{+	0.2	}$ &	$	2.637	_{-	0.004	}^{+	0.004	}$ &			 -					&	0.529		&	0.157		&	0.103		&	PCA+HEXTE	\\
85	&	0.85	&	$	1.02	_{-	0.01	}^{+	0.0	}$ &	$	19.0	_{-	0.2	}^{+	0.2	}$ &	$	2.638	_{-	0.004	}^{+	0.004	}$ &			 -					&	0.522		&	0.157		&	0.098		&	PCA+HEXTE	\\
86	&	0.88	&	$	1.024	_{-	0.004	}^{+	0.0	}$ &	$	22.6	_{-	0.2	}^{+	0.2	}$ &	$	2.699	_{-	0.004	}^{+	0.004	}$ &			 -					&	0.538		&	0.141		&	0.142		&	PCA+HEXTE	\\
87	&	1.10	&	$	0.98	_{-	0.02	}^{+	0.07	}$ &	$	30	_{-	4	}^{+	3	}$ &	$	2.63	_{-	0.05	}^{+	0.05	}$ &			 -					&	0.554		&	0.127		&	0.203		&	PCA+HEXTE	\\
88	&	1.05	&	$	0.929	_{-	0.002	}^{+	0.00	}$ &	$	38.2	_{-	0.2	}^{+	0.2	}$ &	$	2.532	_{-	0.003	}^{+	0.003	}$ &			 -					&	0.488		&	0.092		&	0.246		&	PCA+HEXTE	\\
89	&	1.26	&	$	0.922	_{-	0.002	}^{+	0.00	}$ &	$	38.5	_{-	0.2	}^{+	0.2	}$ &	$	2.525	_{-	0.003	}^{+	0.003	}$ &			 -					&	0.462		&	0.085		&	0.240		&	PCA+HEXTE	\\
90	&	0.97	&	$	0.898	_{-	0.002	}^{+	0.00	}$ &	$	39.6	_{-	0.2	}^{+	0.2	}$ &	$	2.641	_{-	0.003	}^{+	0.003	}$ &			 -					&	0.465		&	0.088		&	0.222		&	PCA+HEXTE	\\
91	&	0.98	&			 -				&			 -				&	$	2.88	_{-	0.01	}^{+	0.02	}$ &			 -					&	0.048		&	0.018		&	0.000		&	PCA+HEXTE	\\
92	&	1.32	&			 -				&			 -				&	$	2.53	_{-	0.01	}^{+	0.02	}$ &			 -					&	0.046		&	0.019		&	0.000		&	PCA+HEXTE	\\
93	&	1.23	&			 -				&			 -				&	$	2.46	_{-	0.01	}^{+	0.01	}$ &			 -					&	0.045		&	0.020		&	0.000		&	PCA+HEXTE	\\
94	&	1.28	&			 -				&			 -				&	$	2.45	_{-	0.01	}^{+	0.01	}$ &			 -					&	0.045		&	0.021		&	0.000		&	PCA+HEXTE	\\
95	&	1.30	&			 -				&			 -				&	$	2.38	_{-	0.01	}^{+	0.02	}$ &			 -					&	0.045		&	0.021		&	0.000		&	PCA+HEXTE	\\
96	&	1.16	&			 -				&			 -				&	$	2.45	_{-	0.01	}^{+	0.01	}$ &			 -					&	0.045		&	0.021		&	0.000		&	PCA+HEXTE	\\
97	&	1.25	&			 -				&			 -				&	$	2.52	_{-	0.01	}^{+	0.01	}$ &			 -					&	0.044		&	0.020		&	0.000		&	PCA+HEXTE	\\
98	&	1.97	&			 -				&			 -				&	$	2.45	_{-	0.05	}^{+	0.06	}$ &			 -					&	0.043		&	0.020		&	0.000		&	PCA+HEXTE	\\
99	&	1.47	&			 -				&			 -				&	$	1.81	_{-	0.02	}^{+	0.02	}$ &			 -					&	0.285		&	0.168		&	0.000		&	PCA+HEXTE *	\\
100	&	1.77	&			 -				&			 -				&	$	1.80	_{-	0.02	}^{+	0.02	}$ &			 -					&	0.293		&	0.172		&	0.000		&	PCA+HEXTE *	\\
101	&	1.55	&			 -				&			 -				&	$	1.82	_{-	0.03	}^{+	0.02	}$ &			 -					&	0.300		&	0.175		&	0.000		&	PCA+HEXTE *	\\
102	&	1.86	&			 -				&			 -				&	$	1.82	_{-	0.03	}^{+	0.03	}$ &			 -					&	0.307		&	0.179		&	0.000		&	PCA+HEXTE *	\\
103	&	1.73	&			 -				&			 -				&	$	1.85	_{-	0.03	}^{+	0.02	}$ &			 -					&	0.317		&	0.183		&	0.000		&	PCA+HEXTE *	\\
104	&	1.68	&			 -				&			 -				&	$	1.86	_{-	0.04	}^{+	0.02	}$ &			 -					&	0.317		&	0.183		&	0.000		&	PCA+HEXTE *	\\
105	&	1.24	&			 -				&			 -				&	$	1.81	_{-	0.03	}^{+	0.03	}$ &	$	171.893	_{-	18	}^{+	24	}$	&	0.335		&	0.194		&	0.000		&	PCA+HEXTE *	\\
106	&	1.36	&			 -				&			 -				&	$	1.85	_{-	0.02	}^{+	0.03	}$ &			 -					&	0.340		&	0.194		&	0.000		&	PCA+HEXTE *	\\
107	&	1.08	&			 -				&			 -				&	$	1.92	_{-	0.04	}^{+	0.03	}$ &	$	191.624	_{-	31	}^{+	37	}$	&	0.347		&	0.190		&	0.000		&	PCA+HEXTE *	\\
108	&	1.25	&			 -				&			 -				&	$	2.10	_{-	0.03	}^{+	0.03	}$ &			 -					&	0.367		&	0.184		&	0.000		&	PCA+HEXTE *	\\
109	&	1.50	&			 -				&			 -				&	$	2.20	_{-	0.00	}^{+	0.00	}$ &			 -					&	0.375		&	0.179		&	0.000		&	PCA+HEXTE *	\\
110	&	1.17	&	$	0.7	_{-	0.3	}^{+	0.57	}$ &	$	39	_{-	10	}^{+	330	}$ &	$	2.29	_{-	0.16	}^{+	0.09	}$ &			 -					&	0.405		&	0.160		&	0.043		&	PCA+HEXTE *	\\
111	&	1.10	&	$	0.8	_{-	0.4	}^{+	0.5	}$ &	$	22	_{-	9	}^{+	188	}$ &	$	2.2	_{-	0.2	}^{+	0.2	}$ &			 -					&	0.392		&	0.158		&	0.046		&	PCA+HEXTE *	\\
112	&	1.55	&	$	0.98	_{-	0.07	}^{+	0.22	}$ &	$	20	_{-	5	}^{+	10	}$ &	$	2.1	_{-	0.1	}^{+	0.2	}$ &	$	153.15	_{-	52	}^{+	150	}$	&	0.392		&	0.144		&	0.083		&	PCA+HEXTE *	\\
113	&	1.19	&	$	1.027	_{-	0.004	}^{+	0.004	}$ &	$	20.9	_{-	0.2	}^{+	0.2	}$ &	$	2.185	_{-	0.004	}^{+	0.004	}$ &			 -					&	0.405		&	0.129		&	0.122		&	PCA+HEXTE *	\\
114	&	1.19	&	$	0.9	_{-	0.1	}^{+	0.3	}$ &	$	30	_{-	7	}^{+	15	}$ &	$	2.4	_{-	0.4	}^{+	0.2	}$ &			 -					&	0.429		&	0.132		&	0.137		&	PCA+HEXTE *	\\
115	&	1.40	&	$	1.002	_{-	0.003	}^{+	0.003	}$ &	$	28.9	_{-	0.1	}^{+	0.1	}$ &	$	2.274	_{-	0.004	}^{+	0.004	}$ &			 -					&	0.454		&	0.100		&	0.214		&	PCA+HEXTE *	\\
116	&	1.11	&	$	0.853	_{-	0.002	}^{+	0.002	}$ &	$	42.8	_{-	0.2	}^{+	0.2	}$ &	$	2.457	_{-	0.004	}^{+	0.004	}$ &			 -					&	0.387		&	0.074		&	0.197		&	PCA+HEXTE *	\\
117	&	1.24	&			 -				&			 -				&	$	2.95316	_{-	0.09	}^{+	0.10	}$ &			 -					&	0.032		&	0.012		&	0.000		&	PCA+HEXTE *	\\
\hline																																									
																														
\end{longtable} 
\end{center} 

\twocolumn

\bibliographystyle{mn2e.bst}
\bibliography{biblio.bib} 
\label{lastpage}
\end{document}